\documentclass[12pt,groupaddress,showpacs,floatfix]{revtex4}

\usepackage{graphicx}
\usepackage{dcolumn}
\usepackage{bm}
\usepackage{amsmath}
\usepackage{amssymb}
\usepackage{times}
\usepackage{amsmath}

\begin{document}

\title{Quantum adiabatic optimization and combinatorial landscapes}

\author{V. N. Smelyanskiy}
\email{Vadim.N.Smelyanskiy@nasa.gov}

\author{S. Knysh}
\email{knysh@email.arc.nasa.gov}

\author{R.D. Morris}
\email{rdm@email.arc.nasa.gov}

\affiliation{NASA Ames Research Center, MS 269-3, Moffett Field,
CA 94035-1000}

\date{\today}

\begin{abstract}
In this paper we analyze the performance of the Quantum Adiabatic
Evolution algorithm on a variant of Satisfiability problem
for an ensemble of random graphs parametrized by the ratio of
clauses to variables, $\gamma=M/N$. We introduce a set of
macroscopic parameters (landscapes) and put forward an ansatz of
universality for random bit flips. We then formulate the problem
of finding the smallest eigenvalue and the excitation gap as a
statistical mechanics problem. We use the so-called annealing
approximation with a refinement that a finite set of macroscopic
variables ({\em versus} only energy) is used, and are able to show
the existence of a dynamic threshold $\gamma=\gamma_d$ starting with 
some value of $K$ -- the number of variables in each clause. 
Beyond dynamic threshold, the algorithm should take exponentially long 
time to  find a solution. We compare the results for extended and 
simplified sets of landscapes and provide numerical evidence in support 
of our universality ansatz. We have been able to map the ensemble of 
random graphs onto another ensemble with fluctuations significantly 
reduced. This enabled us to obtain tight upper bounds on satisfiability
transition and to recompute the dynamical transition using the extended 
set of landscapes.
\end{abstract}

  \maketitle

\section{\label{sec:Intro} Introduction}

An important open question in the field of quantum computing is
whether  it is possible  to develop quantum algorithms capable of
efficiently solving
combinatorial optimization problems (COP). In the simplest case
the task in a COP is to minimize the energy function $E_{\bf
\sigma}$ with the domain given by the set of all possible
assignments of $N$ binary variables, ${\boldsymbol
\sigma}=\{\sigma_1,\ldots, \sigma_N \}$,  $\sigma_j=\pm 1$. In
quantum computation this cost function corresponds to a
Hamiltonian ${\cal H}_P$
\begin{eqnarray}
&& {\cal H}_P=\sum_{\boldsymbol \sigma} E_{\boldsymbol
\sigma} | {\boldsymbol \sigma}\rangle \langle {\boldsymbol \sigma}| \label{HP} \\
&& |{\boldsymbol \sigma}\rangle = | \sigma_1\rangle_1\,
\otimes|\sigma_2\rangle_2\,\otimes\cdots\otimes|\sigma_N\rangle_N
, \nonumber
\end{eqnarray}
\noindent where  the summation is over the $2^N$ states $|{\boldsymbol
\sigma}\rangle$  forming the computational basis of a quantum
computer with $N$ qubits. State $|\sigma_j\rangle_j$ of the $j$-th
qubit is an eigenstate of the Pauli matrix $\hat{\sigma}_z$ with
eigenvalue $\sigma_j$. It is clear from the above that the ground
state of $H_P$ encodes the solution to the COP with cost function
$E_{\boldsymbol \sigma}$. In what follows we shall use two
equivalent notations for binary variables:  Ising spins
$\sigma_j=\pm 1$ as well as bits $z_j=(1-\sigma_j)/2=0,1$.

Recently Farhi and coworkers proposed a new family of quantum
algorithms for combinatorial optimization that is based on the
properties of quantum adiabatic evolution \cite{Farhi,Farhi:Sc}.
Numerical simulations were performed for the study of its performance
for satisfiability problems \cite{Hogg:02}.
Implementation of these algorithms on a quantum computing device
is feasible for COPs  where the energy function $E_{\boldsymbol
\sigma}$ possesses a locality property, in a sense that it is
given by the sum of terms each involving only a relatively small
number of bits, that does not scale with $N$
\cite{Lloyd,Farhi,Kaminsky}. An example of a problem that can have
this property is Satisfiability that deals with $N$ binary
variables, submitted to $M$ constraints, assuming that each
constraint involves ${\cal O}(1)$ bits. The task is to find a bit
assignment that satisfies all the constraints.

Satisfiability is a basic problem in the so-called NP-complete
class \cite{Karp}. This class contains hundreds of the most common
computationally hard problems encountered in practice, such as
constraint satisfaction and graph coloring. NP-complete
problems are characterized in the worst case by exponential
scaling of the run time or memory requirement with the
problem size $N$. A special property of the class is that any
NP-complete problem can be converted into any other NP-complete
problem in polynomial time on a classical computer. Therefore, it
is sufficient to find a deterministic algorithm that can be
guaranteed to solve all instances of just one of the NP-complete
problems within a polynomial time bound. It is widely believed,
however, that such an algorithm  does not exist on a classical
computer. Whether it exists on a quantum computer is one of the
central open questions.

Running of the quantum adiabatic evolution algorithms (QAA) for
several NP-complete problems  has been simulated on a classical
computer using a large number of randomly generated problem
instances that are believed to be computationally hard for
classical algorithms \cite{Farhi:Sc,Farhi:Cli,Hogg:02}. Results of
these numerical simulations for relatively small size of the
problem  instances ( $N \lesssim$ 25) suggest a {\em quadratic}
scaling law of the run time of the QAA  with $N$.

A particularly simple version of Satisfiability   is the NP-complete
Exact Cover problem that was used in  \cite{Farhi:Sc} to study the
performance of QAA. In this problem each constraint is a clause
that involves a subset of $K=3$ binary variables. A given
constraint is satisfied if exactly one of its bits equals 1 and
the rest of the bits equal 0. In the optimization version of this
problem one minimizes the energy function $E_{\boldsymbol \sigma}$
that is equal to the number of constraints violated by a given
bit-assignment ${\boldsymbol \sigma}$. A generalization of this
problem to an arbitrary number $K$ can be called positive 1-in-K
SAT \cite{Moore}.

In practice algorithms for NP-complete problems are characterized
by a wide range of running times, from linear to exponential,
depending on the choice of  certain control parameters of the
problem (e.g., in Satisfiability it is the ratio of the number of
constraints to the number of variables, $M/N$). Therefore, a
practically important alternative to the worst case complexity
analysis is study of a typical-case behavior of optimization
algorithms on ensembles  of randomly generated  problem instances
chosen from a given probability distribution. For example, in the
case of  positive 1-in-K SAT one can define a uniform ensemble of random
problem instances. An instance ${\cal I}$ consists of $M$ statistically
independent clauses, each corresponding to a
$K$-tuple of distinct bit-indices  uniformly  sampled from the interval $(1,N)$
with probability $1/\binom{N}{K}$.

\begin{figure}[bht]\hspace{-0.33in}
\includegraphics[width=3.3in]{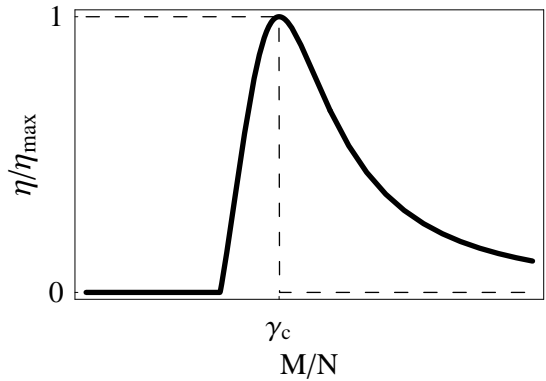}
\caption{\label{fig:complexity} Solid line shows the qualitative
plot of the normalized quantity $\eta/\eta_{\rm max}$ {\it vs}
$M/N$ ($\eta_{\rm max}$ is a maximum value of $\eta$). Dashed line
shows the proportion  of satisfiable instances {\it vs} $M/N$.}
\end{figure}

In the case of an exponential scaling low for the algorithm's running
times  $t_a$ it is convenient  to analyze the distribution of a
normalized logarithmic quantity $\log t_a/N$. This
 distribution
 becomes increasingly narrow in the limit of large $N$ where the mean
 value $\langle \log t_a \rangle/N$ well characterizes  the  typical case
 exponential complexity of an algorithm. For Satisfiability problem
the dependence of the asymptotic quantity
\begin{equation}
\eta=\lim_{N\rightarrow\infty} \langle \log t_a
\rangle/N\label{gamma}
\end{equation}
\noindent  on the clause-to-variable ratio $\gamma=M/N$ has the
qualitative form shown in Fig.\ref{fig:complexity}. At some
critical value $\gamma=\gamma_d$ algorithmic complexity undergoes
the dynamical transition from polynomial to exponential scaling
law. This transition has been studied recently for the case of
a variant of the classical random-walk algorithm for the Satisfiability problem
\cite{Semerjian:03}.
\noindent Function $\eta(\gamma)$ is non-monotonic in $\gamma$
and reaches its maximum at a certain point $\gamma_c>\gamma_d$. It
was discovered some time ago
\cite{Cheeseman:91,Kirkpatrick:94,AI:96} that $\gamma_c$ is a
critical value for the so called satisfiability phase transition:
if $\gamma<\gamma_c$, a randomly drawn instance is satisfiable
with high probability, i.e., there exists at least one bit
assignment ${\boldsymbol \sigma}$ that satisfies all the
constraints ($E_{\boldsymbol \sigma}=0$). For $\gamma> \gamma_c$
instances are almost never satisfiable. In the asymptotic limit
$N\rightarrow \infty$ the proportion of satisfiable instances
drops from 1 to 0 infinitely steeply at $\gamma=\gamma_c$ as shown
in Fig.~\ref{fig:complexity}.

The value of $\gamma_d$ (unlike $\gamma_c$)  depends on both the
problem at hand and the optimization algorithm.
Recent years have seen a rising interest in study of dynamic
threshold phenomena for local search algorithms \cite{Semerjian:03,Barthel:03}.
That effort is
in its initial stage and simple approximations (in spirit of
annealing approximation) were employed to estimate the location
of threshold.
Comparison of the
dynamical thresholds $\gamma_d$ for different algorithms provides
an important relative measure of their typical-case performance in
a given problem. 

This paper is organized as follows. In section \ref{sec:QAA} we introduce
the Quantum Adiabatic Evolution Algorithm and explain how the complexity
of the algorithm depends on the spectrum of the Hamiltonian. In section 
\ref{sec:WKB} we formulate quasiclassical approximation used to study the
complexity and introduce the notion of landscapes. In section \ref{sec:Models}
we introduce positive $K$-NAE SAT and positive $1$-in-$K$ SAT -- the 
NP-complete problems, which we use as a test bed for our method. 
In section \ref{sec:AA} we provide detailed computation of entropy and 
landscapes within annealing approximation. We discuss the universality
of landscape probability distributions in section \ref{sec:Universality}.
Sections \ref{sec:Core} and \ref{sec:CoreQAA} are devoted to improving the 
annealing bound. A subgraph responsible for the hardest part of the problem 
(a core) is identified and results are rederived for the subgraph. In all 
cases we are concerned in finding the dynamic threshold -- the critical ratio 
of clauses to variables above which the algorithm is expected to take 
exponentially long time to find a solution. We discuss our results as well 
as possible ramifications and extensions of our work in Conclusion 
(section \ref{sec:Conclusion}). Appendix \ref{app:np} gives a sketchy
proof of NP-completeness of the problems we considered, and 
appendix \ref{app:next_order} discusses an incremental improvement over 
annealing approximation possible within our formalism.

\section{\label{sec:QAA} Quantum Adiabatic Evolution Algorithm}

Consider the time-dependent Hamiltonian $ \textsl{ H}(t)\equiv
{\cal H}(t/T)$
\begin{equation}
{\cal H}(\tau) = (1-\tau)\, {\cal H}_B+ \tau\, {\cal H}_P,
\label{H}
\end{equation}
\noindent where $\tau=t/T \in (0,1)$ is dimensionless \lq\lq
time", ${\cal H}_P$ is the \lq\lq problem" Hamiltonian (\ref{HP})
and ${\cal H}_B$ is a \lq\lq driver" Hamiltonian, that is designed
to cause transitions between the eigenstates of ${\cal H}_P$.
Using  dimensionless time and setting $\hbar=1$  the quantum state
evolution obeys the equation, $i \,T\partial
|\Psi(\tau)\rangle/\partial \tau={\cal
H}(\tau)|\Psi(\tau)\rangle$. At the initial moment the quantum
state  $|\Psi(0)\rangle$ is prepared to be the ground state of
${\cal H}(0)={\cal H}_B$. In the simplest case
\begin{equation}
 {\cal H}_B = -\sum_{j=1}^{N} \sigma_{x}^{j},\quad
|\Psi(0)\rangle =2^{-N/2}\sum_{\boldsymbol \sigma}|{\boldsymbol
\sigma}\rangle, \label{HB}
\end{equation}
\noindent  where $\sigma_{x}^{j}$ is a Pauli matrix for $j$-th
qubit.  Consider the instantaneous eigenstates of ${\cal H}(\tau)$
with eigenvalues $\lambda_k(\tau)$ arranged in nondecreasing order
at any value of $\tau\in(0,1)$
\begin{equation}
 {\cal H}(\tau)\, |\phi_k(\tau)\rangle =
\lambda_k(\tau)\,|\phi_k(\tau)\rangle, \label{eigen}
\end{equation}
\noindent
 here $ k=0,1,2,\ldots,2^N-1$. Provided the value of $T$ (the runtime of the algorithm)
 is large enough
 and there is a finite gap for all $\tau\in(0,1)$  between the
 ground and excited state energies,
 $\lambda_1(\tau)-\lambda_0(\tau)>0$,
the  quantum evolution is adiabatic and the state of the system
 $|\Psi(\tau)\rangle$  stays
close to an instantaneous ground state, $|\phi_0(\tau)\rangle$ (up
to a phase factor). The state $|\phi_0(1)\rangle$ coincides with
the ground state  of the problem Hamiltonian ${\cal H}_P$ and,
therefore, a measurement performed on the quantum computer at the
final moment $t=T\, (\tau=1)$ will yield one of the solutions of
COP with large probability.

The standard criterion for adiabatic evolution is usually
formulated in terms of minimum excitation gap between the ground
and first exited states \cite{Messiah}
\begin{equation}
T \gg \frac{{\cal E}}{\Delta\lambda_{\rm min}^{2}},\quad
\Delta\lambda_{\rm min}=\max_{0 \leq \tau \leq
1}\left[\lambda_{1}(\tau)-\lambda_{0}(\tau)\right].\label{gap}
\end{equation}
\noindent Here the quantity ${\cal E}$ is less than the largest
eigenvalue of the operator ${\cal H}_P-{\cal H}_B$
\cite{Farhi:annealing} and scales polynomially with $N$ in the
problems we consider.

\section{\label{sec:WKB} Quasiclassical approximation and combinatorial landscapes}

 In the computational basis  (\ref{HP}) we have
\begin{equation} {\cal H} = \tau \sum_{\boldsymbol \sigma} E_{\boldsymbol \sigma} |{\boldsymbol \sigma}
\rangle \langle {\boldsymbol \sigma}| - (1-\tau)
\,\sum_{{\boldsymbol \sigma},{\boldsymbol
\sigma^{\prime}}}\delta\left[ d({\boldsymbol \sigma},{\boldsymbol
\sigma^{\prime}}), 1\right] |{\boldsymbol \sigma} \rangle \langle
{\boldsymbol \sigma^{\prime}}|,\label{HG}
\end{equation}\noindent
here $\delta[m,n]$ denotes the  Kronecker delta-symbol and the
summation is over the pairs of spin configurations ${\boldsymbol
\sigma}$ and ${\boldsymbol \sigma'}$ that differ by the orientation of
a single spin, $d({\boldsymbol \sigma},{\boldsymbol \sigma'})$=1,
where
\begin{equation}
d({\boldsymbol \sigma},{\boldsymbol
\sigma'})=\frac{1}{2}\sum_{j=1}^{N}|\sigma_j-\sigma_{j}^{\prime}|
,\label{hamming}
\end{equation}
\noindent denotes a so-called Hamming distance between the spin
configurations ${\boldsymbol \sigma}$ and ${\boldsymbol \sigma'}$,
that is the number of spins with opposite orientations.
 Eq.~ (\ref{eigen}) in the computational basis takes form
\begin{equation}
 \lambda(\tau) \phi_{\boldsymbol \sigma}(\tau) = \tau E_{\boldsymbol \sigma}
 \phi_{\boldsymbol \sigma}(\tau) - (1-\tau)
 \sum_{{\boldsymbol \sigma}^{\prime}}\delta\left[ d({\boldsymbol \sigma},{\boldsymbol
 \sigma'}),1\right]\,
 \phi_{{\boldsymbol \sigma'}}(\tau)\label{sch}
\end{equation}
\noindent
(here we drop the subscript indicating the number of a quantum
state in $\lambda$ and $\phi_{\boldsymbol \sigma}$). In what
follows we assume that typical energies $E_{\boldsymbol
\sigma}={\cal O}(N)$, but the change in the energy after a single
spin  flip  is ${\cal O}(1)$. This assumption about the energy
landscape holds for instances of the Satisfiability problem with
the clause-to-variable ratio $M/N={\cal O}(1)$, the case of most
interest for us (see the discussion in Sec.~\ref{sec:Intro}).

We now consider a set of functions $\{X_l=C_{l}({\boldsymbol
\sigma},{\cal I}),\,l=1,\ldots,{\cal K}\}$, referred to as
(combinatorial) landscapes, that depend on a problem instance
${\cal I}$  and project a spin configuration ${\boldsymbol
\sigma}$ onto a vector $\{X_l\}$ with integer-valued components.
Prior to considering a specific COP here we make certain assumptions
about the  properties of landscapes and apply them to the analysis
of the minimum gap in the QAA.

In particular, we assume that, similar to energy,
 landscapes $\{X_l=C_l({\boldsymbol \sigma},{\cal I})\}$ are macroscopic
functions, so that  the typical values of $X_l$ are ${\cal O}(N)$, and
possess a certain \emph{universality} property in the asymptotic
limit $N\rightarrow \infty$. Specifically, the joint distribution
of $\{ C_l({\boldsymbol \sigma},{\cal I})\}$  over the spin
configurations ${\boldsymbol \sigma}$ forming the 1-spin-flip
neighborhood of an \lq\lq ancestor" configuration ${\boldsymbol
\sigma'}$ depends  on a problem instance  ${\cal I}$  and spin
configuration ${\boldsymbol \sigma'}$ \emph{only} via the set of
parameters $\{X_{l}^{\prime}=C_l({\boldsymbol \sigma'},{\cal
I})\}$. We then define a quantity
\begin{eqnarray}
&&P\left(\{ X_l\}\,|\,\{
X_{l}^{\prime}\}\right)=\frac{1}{N}\,\sum_{d({\boldsymbol
\sigma},{\boldsymbol \sigma'})=1}\prod_{k=1}^{\cal K} \delta\left[
X_l, C_l({\boldsymbol \sigma},{\cal I})\right],\label{cond}
\\
&& X_{l}^{\prime}=C_{l}({\boldsymbol \sigma'},{\cal I}),\nonumber
\end{eqnarray}
\noindent  In effect, the above universality property of
landscapes implies that the set of all possible spin configurations
${\boldsymbol \sigma}$ is divided into \lq\lq boxes" with
coordinates $\{X_l\}$  where $X_l= C_l({\boldsymbol \sigma})$, and
$P\left(\{ X_l\}\,|\,\{ X_{l}^{\prime}\}\right)$ (\ref{cond})
represents the transition probability from  box $\{X_l\}$ to box
$\{X_{l}^{\prime}\}$. In particular, it obeys  Bayes' rule
\begin{equation}
P\left(\{ X_l\}\,|\,\{
X_{l}^{\prime}\}\right)\,\Omega(\{X_{l}^{\prime}\})=P\left(\{
X_{l}^{\prime}\}\,|\,\{
X_{l}\}\right)\,\Omega(\{X_{l}\}),\label{B}
\end{equation}
\noindent where
 $\Omega(\{X_{l}\})$ is the number of different spin
configurations in the box $\{X_l\}$.

We consider  energy to be a smooth function of landscapes
\begin{equation}
E_{\boldsymbol \sigma}=E\left(\{X_l\}\right),\quad X_l\equiv
C_l({\boldsymbol \sigma},{\cal I}),\label{EviaC}
\end{equation}
\noindent so that $|\partial E/\partial X_l|={\cal O}(1)$.
Furthermore, we assume that, on one hand, the change in
$C_l({\boldsymbol \sigma},{\cal I})$ after flipping one spin is
${\cal O}(1)$, for typical problem instances. On the other hand,
we assume that correlation properties in a neighborhood of a box $\{X_l\}$
described by $P\left(\{ X_l\}\,|\,\{ X_{l}^{\prime}\}\right)$ vary
smoothly with box coordinates on a scale $1\lesssim |\delta X_{l}|
\ll N$. Therefore if we write the transition probability in the
form
\begin{equation}
P\left(\{ X_{l}^{\prime}\}\,|\,\{ X_{l}\}\right)=p\left(\{
X_{l}^{\prime}-X_{l}\};\,\{x_l\}\right),\quad \{x_l\equiv
X_l/N\},\label{cond1}
\end{equation}
then  $p\left(\{ k_l\};\{x_l\}\right)$ is a steep function of its
first argument: it decays rapidly in the range $1\lesssim |k_l|\ll
N$ for each $l$-component. However this is a smooth function of
its second argument:  it varies slightly when coordinates $x_l$
change  on a scale $|\delta x_l| \ll 1$.

One can show  that under the above assumptions the quantum amplitudes
$\phi_{\boldsymbol \sigma}$ corresponding to the smallest
eigenvalue depend on the spin configuration ${\boldsymbol \sigma}$
only via the coordinates of this box $\{X_l\}$ to which it belongs.
Then we look for the solution of (\ref{sch}) in the following
form:
\begin{equation}
\phi_{\boldsymbol \sigma}(\tau) = \frac{\varphi
(\{X_l\},\tau)}{\sqrt{\Omega (\{X_l\})}},\quad \{ X_l\equiv
C_{l}({\boldsymbol \sigma},{\cal I})\}.\label{varphi}
\end{equation}
\noindent where $|\varphi(\{X_l\},\tau)|^2$ gives the probability
of finding the system in the box $\{X_l \}$.   Plugging (\ref{varphi})
into (\ref{sch}) and making use of (\ref{B}),(\ref{EviaC}) we
obtain:
\begin{eqnarray}
&&\lambda(\tau) \varphi ({\bf X},\tau)= \tau E ({\bf X}) \varphi
({\bf X},\tau) -
   (1-\tau) N \sum_{{\bf X^{\prime}}} L ({\bf X},{\bf X^{\prime}}\} )\, \varphi
   ({\bf X^{\prime}},\tau ),\label{sch1}\\
&&{ \bf X}\equiv \{X_1,X_2,\ldots,X_{\cal K}\},\label{short}
\end{eqnarray}\noindent
(hereafter we use the above shorthand notation for the set of
landscapes). In (\ref{sch1}) we introduced
\begin{eqnarray}
&& L({\bf X},{\bf X'})=L({\bf X'},{\bf X})=P ({\bf X}' |{\bf X})
   \sqrt{\frac{P({\bf X})}{P({\bf X}' )}},\label{L}\\
   && P({\bf X})=2^{-N}\Omega({\bf X}),\nonumber
     \end{eqnarray}
   \noindent
where $P({\bf X})$ is a probability that a randomly sampled
configuration ${\boldsymbol\sigma}$  belongs to a box  ${\bf X}$.
We shall look for a solution of (\ref{sch1}) in the WKB-like form
\begin{equation}
\varphi ({\bf X},\tau)= \exp\left(- W ({\bf
X},\tau)\right),\label{WKB}
\end{equation}
\noindent  so that
\begin{equation}
\lambda(\tau) =\tau E ({\bf X}) - (1-\tau) N \sum_{{\bf X}'} L
   ({\bf X},{\bf X}' ) e^{ W ({\bf X},\, \tau) -W
   ({\bf X'},\,\tau) }\label{sch2} .\end{equation}
\noindent We now introduce  scaled variables (cf.
(\ref{cond1}))
\begin{equation}
{\bf x}=\frac{\bf X}{N},\quad \Gamma=\frac{1-\tau}{\tau},\quad
g=\frac{\lambda}{\tau N},\label{scaled}\end{equation} \noindent
and also
\begin{equation}
 \emph{w}({\bf x},\Gamma)\equiv \frac{1}{N} W({\bf X},\tau),\quad
\varepsilon({\bf x})\equiv \frac{1}{N}E({\bf X}),\quad s({\bf
x})\equiv\frac{1}{N}\log\Omega({\bf X}),\label{defs}
\end{equation}
\noindent where $s({\bf x})$ is an entropy function. Based on
(\ref{L}) and the properties of the transition probability  (see
Eq.~(\ref{cond1}) and discussion after it) we assume that the sum
over ${\bf X'}$ in (\ref{sch2}) is dominated by terms with $|{\bf
X'}~-~{\bf X}| = {\cal O}(1)$. Then we can use an approximation
\begin{equation}
W({\bf X}',\tau ) - W ({\bf X},\tau)\approx {\boldsymbol \nabla}
\emph{w}
 \cdot ({\bf X}'
-{\bf X})+{\cal O}(1/N),\label{s}
\end{equation}
\noindent where ${\boldsymbol \nabla} \emph{w} \equiv \partial
\emph{w} ({\bf x},\Gamma)/\partial {\bf x}$. Plugging (\ref{s})
into (\ref{sch2}) and making use of
Eqs.~(\ref{cond1}),(\ref{L}),(\ref{scaled}) and (\ref{defs}) we
obtain after some transformations:
\begin{eqnarray}
 g&=&h({\bf x}, {\boldsymbol \nabla} \emph{w}; \Gamma) ,\label{HJ}\\
h({\bf x},{\bf p};\Gamma)&=&\varepsilon({\bf x})-\Gamma\sum_{{\bf
k}}p({\bf k};{\bf x})e^{-{\bf k}\cdot\left({\boldsymbol \nabla}
\emph{s}/2 +{\bf p}\right)}.\nonumber
\end{eqnarray}
\noindent (here ${\boldsymbol \nabla} \emph{s}\equiv \partial
s({\bf x})/\partial {\bf x}$). This is a Hamilton-Jacobi equation
for an auxiliary mechanical system with coordinates ${\bf x}$,
momenta ${\bf p}={\boldsymbol \nabla}{\emph w}$, action
$\emph{w}$, Hamiltonian function $h({\bf x},{\bf p};\Gamma)$ and
energy $g$. Using the symmetry relation
\begin{equation}
p({\bf k};{\bf x}) e^{-{\bf k}\cdot{\boldsymbol \nabla} s
/2}=p(-{\bf k}; {\bf x}) e^{{\bf k}\cdot{\boldsymbol \nabla} s
/2},\label{sym}
\end{equation}
\noindent that follows directly from Eqs.~(\ref{B}) and (\ref{L})
we obtain that the minimum of   $\emph{w}({\bf x},\Gamma)$ over ${\bf
x}$ where ${\boldsymbol \nabla} \emph{w}=0$ necessary corresponds
to the minimum of the functional:
\begin{equation}
f({\bf x},\Gamma)=\varepsilon({\bf x})-\Gamma \ell ({\bf
x}),\label{F}
\end{equation}
\noindent where $f({\bf x},\Gamma) \equiv h({\bf x},{\bf
0},\Gamma)$ and
 \begin{equation} \ell ({\bf x})
 =\widetilde p\left({\boldsymbol \nabla} s
/2; {\bf x}\right),\quad \widetilde p\left({\bf y};{\bf
 x}\right)\equiv\sum_{{\bf k}}p\left({\bf k};{\bf
 x}\right)e^{-{\bf k}\cdot\,{\bf y}}.\label{calL}
 \end{equation}
 \noindent
The summation in (\ref{HJ}) and (\ref{calL}) is over components
$k_l$ of ${\bf k}$ in the range $k_l\in(-\infty,\infty)$. In what
follows, we shall refer to $\widetilde p({\bf y};{\bf x})$ in
(\ref{calL}) as a \lq\lq Laplace transform" of $p({\bf k}; {\bf
x})$.

We note that $\ell ({\bf x})=\sum_{{\bf X'}}L({\bf X'},{\bf X})$
and one can use Bayes rule and  inequality of Cauchy-Bunyakovsky
in (\ref{L}) to show that that the positive-valued function $\ell({\bf
x})$ is bounded from above, $0< \ell({\bf x}) \leq 1$. This shows
that the analysis of the effective potential based on the WKB
approximation (\ref{s}) is self-consistent in the asymptotic limit
$N\rightarrow\infty$.

 It follows from the
above analysis that the ground-state wavefunction $\psi({\bf
x},\Gamma)\equiv \varphi({\bf X},\tau)$ is concentrated in ${\bf
x}$-space near the bottom of the \lq\lq effective potential" given
by the functional $f({\bf x},\Gamma)$, i.e. near the point ${\bf
x}_{*}(\Gamma)$ where  $f({\bf x},\Gamma)$ reaches its minimum. In
this region $S\approx \frac{1}{2}{\bf x}^T \hat {\bf A} \,{\bf
x}$, where matrix  $\hat {\bf A}$ is positive definite, and
according to (\ref{WKB}), the wavefunction has a Gaussian form with
the width $\propto 1/\sqrt{N}$.

The   ground-state energy $g\equiv g(\Gamma)$ is given by the
value of the effective potential $f$ (\ref{F}) at its minimum
\begin{eqnarray}
&&g(\Gamma) =f({\bf x}_{*}(\Gamma),\Gamma),\label{lambda}\\
&& \left. \partial f({\bf x},\Gamma)/\partial{\bf
x}\right|_{_{{\bf x}={\bf x}_{*}(\Gamma)}}=0,\quad f({\bf
x},\Gamma)\geq g(\Gamma)\nonumber.
\end{eqnarray}
\noindent

We note that as $\Gamma \rightarrow$ 0 the shape of the effective
potential $f({\bf x},\Gamma)$ approaches that of the energy
function $\varepsilon({\bf x})$ and therefore its minimum ${\bf
x}_{*}(\Gamma)\rightarrow {\bf x}_0$ where ${\bf x}_0$ is a
minimum of $\varepsilon({\bf x})$. It can be shown
that in this limit the ground-state eigenvalue approaches the
minimum energy value $\varepsilon({\bf x}_0)$ and the eigenvalues
of ${\bf A}^{-1}$ approach zero (and so does the characteristic
width of the wavepacket $\psi({\bf x},\Gamma)$). The spin
configurations that belong to a box  ${\bf x}_0$ in  ${\bf
x}$-space correspond to the solutions of the optimization problem
at hand. It is clear that one of the solutions can be recovered with
high probability after a measurement is performed at the end of
the  \lq\lq quantum annealing" procedure.

\paragraph*{Variational Ansatz:} For cases in which the set of macroscopic
variables $\{X_l\}$ is not sufficient (in statistical sense
(\ref{cond1})) to describe the dynamics of the quantum algorithm,
one can still implement the above procedure  as an
\emph{approximation}, using a variational method. Introducing a
Lagrangian multiplier $\lambda$, one looks for the minimum of the
functional $F(\varphi,\lambda)=\langle\phi|{\cal H}|\phi
\rangle-\lambda (\langle \phi|\phi\rangle-1)$, using a variational
ansatz (\ref{varphi}) for the wavefunction.  The solution of the
variational problem is provided by
Eqs.~(\ref{WKB})-(\ref{lambda}). The  smallest eigenvalue $g$
(\ref{lambda}) corresponds to the value of the Lagrange multiplier
 at the extremum, $\lambda=\tau N g$, and the maximum of the variational
wavefunction corresponds to the minimum of the effective
potential $f$ (\ref{F}).

\subsection{\label{sec:global} Global bifurcations of the effective potential}
However, in the case of a global bifurcation  where the effective
potential  $f( {\bf x},\Gamma )$ possesses degenerate or nearly
degenerate global minima, the answer is modified. If for some
value of $\Gamma=\Gamma_{\ast}$, a global bifurcation occurs, in
our example this would mean that for this value of $\Gamma$, two
values of ${\bf x}$, ${\bf x}^{+}_{\ast}$ and ${\bf x}^{-}_{\ast}$
give a global minimum to $f( {\bf x},\Gamma )$. In such a case,
the smallest eigenvalue is not doubly degenerate; rather an
exponentially small gap $\Delta\lambda_{\rm min}$ between the
ground and first excited state is developed, itself being
proportional to the overlap between two wave-functions, peaked
around ${\bf x}^{+}_{\ast}$ and ${\bf x}^{-}_{\ast}$ respectively.

To estimate the overlap we note that at $\Gamma_{\ast}$ the two
global minima of the effective potential $f({\bf
x},\Gamma_{\ast})$ correspond to the two coexisting fixed points
of the Hamiltonian function in (\ref{HJ}) with zero momentum and
the same values of energy $g$,
\begin{eqnarray}
&&\partial f/\partial {\bf x}=\partial h/\partial {\bf x}  =\partial h/\partial {\bf p}=0 \\
&&{\bf x}={\bf x}_{*}^{\pm},\quad {\bf p}={\bf
p}_{*}^{\pm}=0,\quad g({\bf x},{\bf
p};\Gamma_{\ast})=g_{*}^{+}=g_{*}^{-}.\label{fixed}
\end{eqnarray} \noindent  Then to logarithmic accuracy we have
\begin{equation}
 \frac{1}{N}\log\Delta g_{\rm min}=
\int_{-\infty}^{\infty}dt'\, [\,\dot{\bf x}(t'){\bf p}(t')-h({\bf
x}(t'),{\bf p}(t'))\,]+{\cal O}(1/N),\label{gapWKB}
\end{equation}
\noindent where  $\left(\,{\bf x}(t),{\bf p}(t)\,\right)$ is a
heteroclinic trajectory connecting the two fixed points of
(\ref{HJ})
\begin{eqnarray}
&& \dot {\bf x}(t)=\partial h/\partial{\bf p} ,\quad \dot {\bf
p}(t)=-\partial h/\partial {\bf x},\label{traj}\\
&& {\bf x}(t\rightarrow \pm\infty)={\bf x}_{*}^{\pm},\quad {\bf
p}(t\rightarrow \pm\infty)=0.\nonumber\label{fixed2}
\end{eqnarray}
\noindent

 From the algorithmic perspective this means that when
$\Gamma$ gets close to $\Gamma_{\ast}$, it has to change
exponentially slowly (cf. Sec. \ref{sec:QAA} and Eq.~(\ref{gap})).
This could be called a critical slowing down in the vicinity of a
quantum phase transition. If  simulated annealing (SA) is used and
a similar phenomenon occurs, the value of the temperature
$T_{\ast}$ is the point where a global bifurcation occurs in the
free energy functional
\begin{equation}
f ({\bf x},T) = \varepsilon ({\bf x}) - T s ({\bf x}).\label{SA}
\end{equation} \noindent  By
comparing the free energy functional (\ref{SA}) with the
 functional  (\ref{F}) corresponding to ``quantum
annealing'' (QA), we note that in QA the quantities  $\Gamma$ and $\ell
({\bf x})$  play the roles of  temperature and entropy in (SA),
respectively.

We note in passing that a similar picture for the onset of global
bifurcation that can lead to the failure of QA and (or) SA was
proposed in \cite{Farhi:annealing,Vazirani:02}  for the case where
the energy
 $E_{\boldsymbol \sigma}$  is a non-monotonic function of  a single
landscape parameter, a total spin $\sum_{j=1}^{N}\sigma_{j}$. In
this case the dynamics of QA  can be described  in terms of
one-dimensional effective potential \cite{Farhi:paths,BS:paths}.

\section{\label{sec:Models}The Models}

 An  instance of a Satisfiability problem with $N$
binary variables committed to $M=\gamma N$ constraints (where each
constraint is a clause involving $K$ variables) can be defined by
the specification of the following two objects. One of them is an
$M\times N$ matrix $\hat {\cal G}$, the rows of the matrix are
independent $K$-tuples of distinct bit indexes  sampled from the
interval $(1,N)$.  The $m\textrm{-th}$ row of $\hat {\cal G}$
defines the subset of the $K$ binary variables involved in the
$m\textrm{-th}$ clause. The second object is a set of boolean
functions ${\cal B}=\{\textsl{b}_m\}$, with each function encoding
a corresponding constraint. A function
$\textsl{b}_m=\textsl{b}_m[\sigma_{_{{\cal G}_{m
1}}},\sigma_{_{{\cal G}_{m 2}}},\ldots, \sigma_{_{{\cal G}_{m
K}}}]$ is defined over the set of $2^K$ possible assignments of
the string of $K$ binary variables  involved in the $m$-th clause.
The function returns value 1 for  assignments of binary variables
that satisfy the constraint and 0 for bit assignments that violate
it. Then the energy function equals to the number of violated
constraints
\begin{equation}
E_{\boldsymbol \sigma}\equiv E_{\boldsymbol \sigma}({\cal
I})=M-\sum_{m=0}^{M}\textsl{b}_m[\sigma_{_{{\cal G}_{m
1}}},\sigma_{_{{\cal G}_{m 2}}},\ldots, \sigma_{_{{\cal G}_{m
K}}}]\label{cost},
\end{equation}
\noindent here ${\cal I}=({\cal G},{\cal B})$ denotes an instance
of a problem.

The matrix $\hat {\cal G}$ defines a hypergraph ${\cal G}$ that is
made up of the set of $N$ vertices (corresponding to the variables in
the problem) and a set of $M$ hyperedges (corresponding to the
constraints of the problem), each one connecting $K$ vertices.
 An ensemble of  \emph{disorder
configurations} of the hypergraph corresponds to all the possible
ways one can place $M=\gamma N$ hyperedges among $N$ vertices
where each hyperedges carries $K$ vertices. Under the uniformity
ansatz all configurations of disorder are sampled with equal
probabilities (i.e.,  rows of the matrix $\hat {\cal G}$ are
independently and uniformly sampled in the (1,$N$) interval).

Boolean functions $b_m$ may also be generated at random for each
constraint with an example being random K-SAT problem
\cite{Monasson:prl96,Monasson:pre97}. However here we consider
slightly different  versions of the random Satisfiability problem
that are still defined on a random hypergraph ${\cal G}$ but have
a non-random boolean function $\textsl{b}_m=\textsl{b}$, identical
for all the clauses in a problem.  One of the problems is Positive
1-in-K Sat
  in which a
constrain is satisfied if and only if exactly one bit is equal 1
and the other $K$-1 bits are equal 0. The boolean function
$\textsl{b}$ for this problem takes the form
\begin{eqnarray}
&&\textsl{b}[\alpha_1,\alpha_2,\ldots,\alpha_K]=\delta\left[\sum_{p=1}^{K}\frac{1-\alpha_p}{2},1\right]
\quad  (\textrm{Positive 1-in-K SAT}) .\label{1inK}\\
&&\alpha_{p}=\pm 1,\quad p=1,2,\ldots,K.\nonumber
\end{eqnarray}
\noindent We shall also consider another problem, Positive
K-NAE-Sat, in which a clause is satisfied unless all variables
that appear in a clause are equal ("K-Not-All-Equal-Sat").
 The boolean function $\textsl{b}$ for this problem takes the form
\begin{equation}
\textsl{b}[\alpha_1,\alpha_2,\ldots,\alpha_K]=
 1-\sum_{s=\pm
 1}\delta\left[\sum_{p=1}^{K}\frac{1+s\,\alpha_p}{2},\,0\right]
\quad (\textrm{Positive K-NAE-SAT})  .\label{nae}
 \end{equation}
Both problems are NP-complete  (Appendix \ref{app:np}). It will be
shown  below that they are characterized by the same set of
landscape functions.

\section{\label{sec:AA}Landscapes: Annealing Approximation}

For a particular spin (${\boldsymbol \sigma})$ and disorder
(${\cal G})$)  configurations, all clauses can be divided into
$2^K$ distinct groups according to the values of the binary variables
that appear in a clause. We will label the different types of
clauses by vectorial index ${\boldsymbol
\alpha}=\{\alpha_1,\ldots,\alpha_K\},\, \alpha_p = \pm 1$. We now
divide the set of $2^N$ spin configurations into boxes identified
by  certain numbers of clauses of each type, $N\,{\cal M}_{\boldsymbol
\alpha}$, and also by  the Ising spin  in a configuration $N q$
\begin{eqnarray}
&&{\cal M}_{\boldsymbol \alpha}\equiv {\cal M}_{\boldsymbol \alpha}({\boldsymbol
\sigma},{\cal
G})=\frac{1}{N}\sum_{m=1}^{M}\prod_{p=1}^{K}\delta\left[
\sigma_{_{{\cal G}_{m p}}},\,\alpha_{p}\right],\label{Mdef}\\
&& q\equiv
q({\boldsymbol\sigma})=\frac{1}{N}\sum_{j=1}^{N}\sigma_j.\label{Sdef}
\end{eqnarray}
\noindent Different boxes correspond to macroscopic states defined
by the set of parameters ($q$, $\{{\cal M}_{\boldsymbol\alpha}\}$) with
$q\in(-1,1)$ and $\sum_{\boldsymbol\alpha}{\cal M}_{\boldsymbol
\alpha}=\gamma$. The energy function can be expressed via (\ref{Mdef})
as follows (cf. (\ref{cost})-(\ref{nae})):
\begin{equation}
\varepsilon\left(\{{\cal M}_{\boldsymbol\alpha}\}\right)=\gamma-\sum_{m=0}^{K}\zeta_{m}
M_{m},\quad M_{m}\equiv\sum_{\boldsymbol\alpha}{\cal M}_{\boldsymbol
\alpha}\,\delta\left[K-2m,\,\sum_{p=1}^{K}\alpha_p\right],\label{eM}
\end{equation}
\noindent where the form of the coefficients $\zeta_m$ depends on
the problem:
\begin{equation}
\zeta_m=\left[\begin{array}{cc}
 \delta[m,1] & \textrm{(Positive 1-in-K SAT)} \\
 1-\delta[m,0]-\delta[m,K] & \textrm{(Positive K-NAE-SAT)} \\
\end{array}\right.\label{zeta}
\end{equation}
\noindent

In the following we compute an approximation to the effective
potential (\ref{F}), using the landscape functions (\ref{Mdef}),
(\ref{Sdef}). According to (\ref{calL}) it depends on the entropy
function $s(q,\{{\cal M}_{\boldsymbol\alpha}\})$ and the transition
probability (\ref{cond1}) between different macroscopic states.
Recalling that variables $q$ and ${\cal M}_{\boldsymbol\alpha}$ are
normalized by the factor $N$ we study the probability of transition,
$p(n,\{r_{\boldsymbol \alpha}\}; q,\{{\cal M}_{\boldsymbol \alpha}\})$,
 from the state  $\left(q,\{{\cal M}_{\boldsymbol \alpha}\}\right)$
 to the  state $\left(q+n/N,\{{\cal M}_{\boldsymbol \alpha}+
r_{\boldsymbol \alpha}/N\}\right )$. The Laplace transform of $p$
with respect to $n,\{r_{\boldsymbol \alpha}\}$  has the form (cf.
(\ref{calL}))
\begin{equation}
\widetilde p\left(\theta,\{y_{\boldsymbol \alpha}\};
q,\{{\cal M}_{\boldsymbol \alpha}\}\right)=\sum_{n,\,\{r_{\boldsymbol
\alpha}\}}e^{-\theta n-\sum_{\boldsymbol \alpha}y_{\boldsymbol
\alpha} r_{\boldsymbol \alpha}} p(n,\{r_{\boldsymbol \alpha}\};
q,\{{\cal M}_{\boldsymbol \alpha}\}),\label{Laplace_def}
\end{equation}
\noindent We assume that all binary variables are also subdivided
into distinct groups based on their value $\sigma= \pm 1$ and a
vector ${\bf k}$ with integer coefficients $k_{\boldsymbol
\alpha}^p$ indicating the number of times a variable appears in a
clause of type ${\boldsymbol \alpha}$ in position $p$. Clearly,
consistency requires that $k_{\boldsymbol \alpha}^p = 0$ unless
$\alpha_p = \sigma$.  We now define a quantity $c_{\sigma,{\bf
k}}$ which is equal to the  fraction of spins with given
$\sigma,{\bf k}$. For a  spin configuration ${\boldsymbol \sigma}$
there exists a set  of coefficients $\{c_{\sigma,{\bf k}}\}$ with
elements of the set corresponding to all possible values of $\sigma$
and ${\bf k}$ (there will be many $0$'s in a set for each spin
configuration).  In general, there are exponentially many sets
$\{c_{\sigma,{\bf k}}\}$ that correspond to a  macroscopic state
($q$, $\{{\cal M}_{\boldsymbol\alpha}\}$)
\begin{equation}
\sum_{\sigma,{\bf k}} \sigma \,c_{\sigma,{\bf k}} = q,\quad
\sum_{\sigma,{\bf k}} k_{{\boldsymbol \alpha}}^p\, c_{\sigma,{\bf
k}} = {\cal M}_{{\boldsymbol \alpha}}\quad
(p=0,1,\ldots,K).\label{consistency}
\end{equation}
\noindent Coefficients $\{c_{\sigma,{\bf k}}\}$ are concentrations
of spin variables  with different types of \lq\lq neighborhoods".
We shall assume that in the limit of large $N$ the distribution of
coefficients $c_{\sigma,{\bf k}}$ corresponding to the same
macroscopic state (\ref{consistency}) is sharply peaked around
their mean values (with the width of the distribution $\propto
N^{-1/2}$).

Under the above assumption we can immediately compute the
Laplace-transformed transition probability (\ref{Laplace_def}) in
terms of the coefficients $c_{\sigma,{\bf k}}$. Indeed, consider
flipping a spin with value $\sigma$ and neighborhood type given by
vector  ${\bf k}$. This will  change the total spin by $- 2
\sigma$ and for each clause of type ${\boldsymbol \alpha}$ and index
$p\in (1,K)$ the value of $N {\cal M}_{{\boldsymbol \alpha}}$ will
decrease by $k_{{\boldsymbol \alpha}}^p$. On the other hand, for
the clause type ${\boldsymbol \alpha}^{\prime}\equiv
\bar{{\boldsymbol \alpha}}(p,{\boldsymbol \alpha})$  obtained by
flipping a bit in $p$-th position in ${\boldsymbol \alpha}$, $N
{\cal M}_{{\boldsymbol \alpha}'}$ is correspondingly increased by
$k_{{\boldsymbol \alpha}}^p$. Hence the Laplace-transformed
transition probability is
\begin{equation}
\widetilde p\left(\theta,\{y_{\boldsymbol \alpha}\};
q,\{{\cal M}_{\boldsymbol \alpha}\}\right)=
   \sum_{\sigma,{\bf k}} c_{\sigma,{\bf k}} \exp\left[2 \theta\sigma +
   \sum_{p,{\boldsymbol \alpha}}\left(y_{{\boldsymbol \alpha}}-y_{\bar{{\boldsymbol \alpha}}
   (p,{\boldsymbol \alpha})}\right)k_{{\boldsymbol
   \alpha}}^p\right] .\label{Laplace} \end{equation}
where the coefficients $c_{\sigma,\,{\bf k}}$ are set to their mean
values in a macroscopic state (\ref{consistency})).

\subsection{\label{sec:entropy} Entropy and coefficients
$c_{\sigma,{\bf k}}$ in a macroscopic state defined by  $q$ and
$\{ {\cal M}_{{\boldsymbol \alpha}}\}$}

Here  we  use  the annealing approximation to estimate the mean values
of $c_{\sigma,\,{\bf k}}$ and also of  a macroscopic state
$(q,{\cal M}_{\boldsymbol\alpha})$. We start by introducing the concept
of annealed entropy. Let $\mathcal{N}$ be the number of spin
configurations subject to some constraints. In general, it is a
function of the disorder realization. The annealed entropy is defined
as the logarithm of its disorder average: $s_{\textrm{ann}} = \ln
\langle \mathcal{N} \rangle$. Note that for the correct, quenched,
entropy the order of taking a logarithm and disorder average is
reversed.

Since in the  random hypergraph model all disorder configurations
are equally probable,  annealed entropy is given as $s_{\rm ann} =
\ln \mathcal{N}_{S, {\cal G}} - \ln \mathcal{N}_{\cal G}$, where
$\mathcal{N}_{S, {\cal G}}$ is the total number of spin and
disorder configurations and $\mathcal{N}_{\cal G}$ is the number
of disorder configurations.

For enumerating all possible disorder configurations we depart slightly
 from the traditional random hypergraph model. In our model
all clauses are ordered (two disorder configurations where any two
clauses are permuted are deemed different), clauses can be
repeated (the same clause can appear twice), the order of variables in
a clause is important (two disorder configurations are different
if the order of variables in any clause is changed), and finally,
variables can be repeated in a single clause. This change does not
alter the underlying physics, since the probability that two
identical clauses appear is infinitesimal, and a variable enters a
clause twice in at most $O ( 1 )$ clauses, which can be safely
neglected. As regards the distinction between the disorders with
permuted clauses, this only introduces a combinatorial factor
which cancels out. The advantage is that each disorder  can be
represented as a sequence of $M$  $K$-tuples of integers from
$1$ to $N$.

We will first compute the annealed entropy of a macroscopic state
$(q,\{{\cal M}_{\boldsymbol \alpha}\})$ under  additional
constraints: we  fix the values $c_{\sigma,{\bf k}}$ and compute
the annealed entropy as a function of $q, \{ {\cal M}_{\boldsymbol
\alpha} \}, \{ c_{\sigma,{\bf k}} \}$. Recalling that
${\cal M}_{\boldsymbol\alpha}$ are the numbers of clauses of a given type
scaled by $N$, and the total number of clauses is $\gamma N$, we
obtain the number of joint spin-disorder configurations as a product
of the following factors:
\begin{itemize}
    \item[(i)]the number of ways to assign types to clauses $( N \gamma ) ! /
\prod_{{\boldsymbol \alpha}}( N {\cal M}_{{\boldsymbol \alpha}} ) !$,
    \item[(ii)]the number of ways to assign types to variables $N! /
\prod_{\sigma,{\bf k}} ( N c_{\sigma,{\bf k}} ) !$,
    \item[(iii)]for all $p,{\boldsymbol \alpha}$, the number of
ways to permute the appearance of variables in $p$-th position of
clauses of type ${\boldsymbol \alpha}$: $(N {\cal M}_{{\boldsymbol
\alpha}}) ! / \prod_{\sigma,{\bf k}} ( k_{{\boldsymbol \alpha}}^p
! )^{N c_{\sigma,{\bf k}}}$,
\end{itemize}
 Consequently, the annealed entropy is given by
\begin{equation} s_{\rm ann} [ \{ c_{\sigma,{\bf k}} \} ; q, \{
{\cal M}_{{\boldsymbol \alpha}} \}
   ] = - \sum_{\sigma,{\bf k}} c_{\sigma,{\bf k}} \ln \left[
   c_{\sigma,{\bf k}} \prod_{p,{\boldsymbol \alpha}} ( k_{{\boldsymbol \alpha}}^p !
   ) \right] + ( K - 1 ) \sum_{{\boldsymbol \alpha}} {\cal M}_{{\boldsymbol \alpha}} \ln
   {\cal M}_{{\boldsymbol \alpha}} + \gamma \ln \gamma -  \gamma K. \label{Hann1}\end{equation}
   \noindent
In the large $N$ limit we replace $c_{\sigma,{\bf k}}$ by their
annealed averages, i.e., the values that maximize the annealed
entropy. In its simplest form, we place no constraints on
$c_{\sigma,{\bf k}}$ except consistency requirements
(\ref{consistency}). Associating Lagrange multipliers $\lambda$
and $\ln \mu_{\boldsymbol \alpha}^p$ with these constraints, the
expression for the entropy can be rewritten as
\begin{eqnarray}
 s_{\rm ann} [ q, \{ {\cal M}_{{\boldsymbol \alpha}} \} ] &=&
\min_{\lambda,\,\,
   \mu_{{\boldsymbol \alpha}}^p} \left\{ - \lambda q + \sum_{p,\,{\boldsymbol \alpha}}
   {\cal M}_{{\boldsymbol \alpha}} \ln
   \frac{{\cal M}_{{\boldsymbol \alpha}}}{\mu_{{\boldsymbol \alpha}}^p} + \ln Z [ \lambda,
   \{ \mu_{{\boldsymbol \alpha}}^p \} ] \right\} \nonumber \\
    &-&\sum_{{\boldsymbol \alpha}} {\cal M}_{{\boldsymbol \alpha}} \ln {\cal M}_{{\boldsymbol \alpha}}+\gamma \ln \gamma -
   \gamma K. \label{Hann2}
   \end{eqnarray}
The values of $c_{\sigma,{\bf k}}$ are given by
\begin{equation}
 c_{\sigma,{\bf
k}} = \frac{1}{Z} e^{\lambda \sigma} \prod_{p,\,{\boldsymbol
\alpha}}
   ( \mu_{{\boldsymbol \alpha}}^p )^{k_{{\boldsymbol \alpha}}^p} /
   k_{{\boldsymbol \alpha}}^p !,\label{c} \end{equation}
and $Z$ is given by
\begin{equation} Z = \exp \left( \lambda +
\sum_{{\boldsymbol \alpha}}\sum_{p}\delta[\alpha_p,1]
   \mu_{{\boldsymbol \alpha}}^p \right) + \exp \left( - \lambda + \sum_{{\boldsymbol \alpha}}
   \sum_{p}\delta[\alpha_p,-1]\mu_{{\boldsymbol \alpha}}^p \right)
   .\label{Z}
\end{equation}
The  values of the Lagrange multipliers $\lambda$, $\mu_{{\boldsymbol
\alpha}}^p$ are related to $q$, $\{ {\cal M}_{{\boldsymbol \alpha}} \}$
via
\begin{eqnarray}
  \frac{\partial \ln Z}{\partial \lambda} & = & q,\label{S1}\\
  \mu_{{\boldsymbol \alpha}}^p  \frac{\partial \ln Z}{\partial
  \mu_{{\boldsymbol \alpha}}^p} & = & {\cal M}_{{\boldsymbol \alpha}}
  .\label{M1}
\end{eqnarray}
\noindent From here we obtain the expression for the Lagrange
multiplier $ \mu_{{\boldsymbol \alpha}}^p$
\begin{equation}
\frac{{\cal M}_{{\boldsymbol \alpha}} }{ \mu_{{\boldsymbol \alpha}}^p} =
\frac{1 + \alpha_p\, q}{2}.\label{mu}
\end{equation}
\noindent Then introducing a new notation
\begin{eqnarray}
  \mu_{\pm} & = & \sum_{p,\,{\boldsymbol \alpha}} \frac{1 \pm \alpha_p}{2}
  \mu_{{\boldsymbol \alpha}}^p,\nonumber\\
  {\cal M}_{\pm} & = & \sum_{p,\,{\boldsymbol \alpha}} \frac{1 \pm \alpha_p}{2}
  {\cal M}_{{\boldsymbol \alpha}}, \label{notation}
\end{eqnarray}
we obtain
\begin{equation} Z = e^{\lambda}
e^{\mu_+} + e^{- \lambda} e^{\mu_-},\quad \mu_{\pm} = \frac{2
{\cal M}_{\pm} }{ 1 \pm q }.\label{Znew}\end{equation} \noindent Then the
entropy can be rewritten in the following form
\begin{equation}
 s_{\rm ann} [ q, \{ {\cal M}_{{\boldsymbol \alpha}} \} ] = - \lambda
q + {\cal M}_+ \ln
   \frac{1 + q}{2} + {\cal M}_- \ln \frac{1 - q}{2} + \ln Z  -
   \sum_{{\boldsymbol \alpha}} {\cal M}_{{\boldsymbol \alpha}} \ln {\cal M}_{{\boldsymbol \alpha}} + \gamma \ln \gamma- \gamma K.
   \label{Hann3}
   \end{equation}
   \noindent
We now use the following equations
\begin{equation}
  e^{\lambda} e^{\mu_+}= Z \frac{1 + q}{2},\quad
  e^{- \lambda} e^{\mu_-} = Z \frac{1 - q}{2}
\end{equation}
and obtain the expression for the second Lagrange multiplier
$\lambda$
\begin{equation} - \lambda q = - \frac{1 + q}{2} \ln
\frac{1 + q}{2} - \frac{1 - q}{2} \ln \frac{1 -
   q}{2} - \ln Z + \gamma K.  \label{lambda1}\end{equation}\noindent
Upon substitution of $\lambda$ from the above into the expression for
$s_{\rm ann}$ (\ref{Hann3}) we finally obtain the annealed entropy
\begin{eqnarray}
 s_{{\rm ann}} [ q, \{ {\cal M}_{{\boldsymbol \alpha}} \} ] = &-& q\, \tanh^{-1}q - \ln
   \frac{\sqrt{1 - q^2}}{2} + {\cal M}_+ \ln \frac{1 + q}{2} + {\cal M}_- \ln \frac{1 -
   q}{2} \nonumber \\&-& \sum_{{\boldsymbol \alpha}} {\cal M}_{{\boldsymbol \alpha}} \ln
   {\cal M}_{{\boldsymbol \alpha}}+ \gamma \ln \gamma . \label{Hann4}\end{eqnarray}\noindent
Also the coefficients $c_{\sigma,{\bf k}}$ are given by
(\ref{c}),(\ref{Z}) with Lagrange multipliers given in (\ref{mu})
and (\ref{lambda1}).

\subsection{\label{sec:potential} Effective potential}

Consider a factor $\ell({\bf x})=(\partial
f/\partial\Gamma)_{\varepsilon}$ (\ref{F}), (\ref{calL}) in the
expression (\ref{F}) for effective potential with  ${\bf x}\equiv
(q,\{{\cal M}_{\boldsymbol\alpha}\})$. It follows from (\ref{calL}) that
to find this factor we need to evaluate the Laplace-transformed
probability (\ref{Laplace_def},\ref{Laplace})) at
\begin{equation}\theta = \frac{1}{2}
\partial s_{\rm ann} /
\partial q,\quad y_{{\boldsymbol \alpha}} = \frac{1}{2} \partial
s_{\rm ann} /
\partial {\cal M}_{{\boldsymbol \alpha}}.\label{theta}\end{equation}
\noindent This is where the Lagrange multipliers come in handy as we
can immediately claim that
\begin{eqnarray}
  \partial s_{\rm ann} / \partial q & = & - \lambda\\
  \partial s_{\rm ann} / \partial {\cal M}_{{\boldsymbol \alpha}} & = & \sum_p \ln
  \frac{{\cal M}_{{\boldsymbol \alpha}}}{\mu_{{\boldsymbol \alpha}}^p} - \ln
  {\cal M}_{{\boldsymbol \alpha}}.\label{relationS}
\end{eqnarray}
Note that in differentiating with respect to ${\cal M}_{{\boldsymbol
\alpha}}$ above we omitted the constant term. This is permissible
since only differences $\partial q_{{\rm ann}} /
\partial {\cal M}_{{\boldsymbol \alpha}} - \partial q_{{\rm ann}} /
\partial {\cal M}_{{\boldsymbol \alpha}'}$ appear in Eq.~(\ref{Laplace}). A further
refinement is to write
\begin{equation} \sum_p \ln \frac{{\cal M}_{{\boldsymbol \alpha}}}{\mu_{{\boldsymbol \alpha}}^p} =
   \sum_p \ln \frac{1 + \sigma_p \,q}{2} = K \ln \frac{\sqrt{1 - q^2}}{2} +
   \sum_p \sigma_p\, \tanh^{-1} q. \end{equation}
Using this in  the Eqs.~(\ref{calL}),(\ref{Laplace}), we obtain
\begin{equation}
\ell(q,\{{\cal M}_{\boldsymbol\alpha}\}) = \frac{1}{Z}
\sum_{\sigma,{\bf k}} \prod_{p,{\boldsymbol \alpha}} \left(
   \mu_{{\boldsymbol \alpha}}^p e^{\frac{1}{2} \sum_{p'} ( \sigma_{p'} -
   \sigma'_{p'} )\, \tanh^{-1} q}
   \sqrt{\frac{{\cal M}_{{\boldsymbol \alpha}'}}{{\cal M}_{{\boldsymbol \alpha}}}}
   \right)^{k_{{\boldsymbol \alpha}}^p} / k_{{\boldsymbol \alpha}}^p !
   \label{ell0}.\end{equation}\noindent
Since $\frac{1}{2} \sum_{p'} ( \sigma_{p'} - \sigma'_{p'} ) \equiv
\sigma_p$ (where ${\boldsymbol \alpha}'$ is obtained from
${\boldsymbol \alpha}$ by flipping $p$-th bit)  and also
\begin{equation}{\cal M}_{{\boldsymbol \alpha}} / \mu_{{\boldsymbol \alpha}}^p =
\frac{\sqrt{1 - q^2}}{2} e^{\sigma_p
\tanh^{-1} q},\end{equation}\noindent the expression is considerably
simplified
\begin{equation}
\ell(q,\{{\cal M}_{\boldsymbol\alpha}\}) =  \frac{2}{Z} \exp \left(
\frac{2}{\sqrt{1 - q^2}}
   \sum_{<{\boldsymbol \alpha},{\boldsymbol \alpha}' >} \sqrt{{\cal M}_{{\boldsymbol \alpha}}
   {\cal M}_{{\boldsymbol \alpha}'}} \right), \end{equation}\noindent
where the sum is over pairs $\langle {\boldsymbol \alpha}$,
${\boldsymbol \alpha}'\rangle$ that differ in exactly one position
\begin{equation}
\frac{1}{2}\sum_{p=1}^{K}|\alpha_j-\alpha_{j}^{\prime}|=1.\label{hamalpha}
\end{equation}
\noindent
 To evaluate $Z$ we write
\begin{equation} Z = \frac{2}{\sqrt{1 - q^2}}  \sqrt{e^{\mu_+} e^{\mu_-}} = \frac{2}{\sqrt{1
   - q^2}} \exp \left( \frac{{\cal M}_+}{1 + q} + \frac{{\cal M}_-}{1 - q} \right) \end{equation}
and the expression for $\ell$ becomes
\begin{equation}
\ell(q,\{{\cal M}_{\boldsymbol\alpha}\}) = \sqrt{1 - q^2} \exp \left(
\frac{2
   \sum_{<{\boldsymbol \alpha},{\boldsymbol \alpha}' >} \sqrt{{\cal M}_{{\boldsymbol \alpha}}
   {\cal M}_{{\boldsymbol \alpha}'}}}{\sqrt{1 - q^2}} - \frac{{\cal M}_+}{1 + q} - \frac{{\cal M}_-}{1
   - q} \right) . \label{ell1}\end{equation}
   \noindent
here ${\cal M}_{\pm}$ are given in (\ref{notation}).

We note that the effective potential
$f(q,\{{\cal M}_{\boldsymbol\alpha}\})=\varepsilon(\{{\cal M}_{\boldsymbol\alpha}\})-\Gamma
\ell(q,\{{\cal M}_{\boldsymbol\alpha}\})$ is symmetric with respect to
permutation of individual components in
$\{{\cal M}_{\boldsymbol\alpha}\}$  corresponding to different orders of
-1's and +1's in the vectorial index ${\boldsymbol \alpha}$. We
look for the minimum of $f(q,\{{\cal M}_{\boldsymbol\alpha}\})$ using
symmetric  ansatz
\begin{equation}
{\cal M}_{\alpha} = \binom{K}{m}_{\overset{}{}}^{-1} M_m ,\quad
m=\sum_{p=1}^{K}\frac{1-\alpha_p}{2}.\label{ansatz}
\end{equation}
\noindent  where $m$ is the number of -1's in ${\boldsymbol
\alpha}$. Substituting (\ref{ansatz}) into (\ref{ell1}) and
rewriting
\begin{eqnarray}
\bar \ell\left(q,\{M_m\}\right)&=& \sqrt{1 - q^2} \exp
\left( \frac{2 \sum_{m = 0}^{K - 1} \sqrt{( m + 1 )
   ( K - m ) M_m M_{m + 1}}}{\sqrt{1 - q^2}}\right.\nonumber \\
    &-&\left. \frac{K \gamma + q \sum_{m = 0}^K (K- 2
   m ) M_m}{1 - q^2} \right).\label{ell2} \end{eqnarray}
   \noindent
where we defined  $\bar\ell\left(q,\{M_m\}\right)\equiv
\ell(q,\{{\cal M}_{\boldsymbol\alpha}\})$. The  effective potential is
then
\begin{equation}
\bar f(q,\{M_m\})=\varepsilon\left(\{M_m\}\right)
-\Gamma \bar \ell\left(q,\{M_m\}\right)\qquad
(QA),\label{FQA}
\end{equation}
\noindent with energy given in (\ref{eM}). In the case of the SA
algorithm the corresponding free-energy functional (\ref{SA}) is
\begin{equation}
\bar f(q,\{M_m\})=\varepsilon(\{M_m\})- T\bar
s(q,\{M_m\})\qquad (SA),\label{FSA}
\end{equation}
\noindent where the entropy function equals
\begin{eqnarray}
\bar s(q,\{M_m\})=&-&q \tanh^{-1} q + (\gamma K-1) \ln
\frac{\sqrt{1 -
   q^2}}{2} \nonumber \\ &-& \left( \sum_{m=0}^{K} ( K-2 m  ) M_m \right) \tanh^{-1} q
   - \sum_{m=0}^{K} M_m
   \ln \frac{M_m}{\binom{K}{m}}.\label{sbar}
\end{eqnarray}
\noindent

 If we were to use an
even smaller set of macroscopic parameters (e.g. only the energy
$\varepsilon $) we can still employ formula (\ref{ell2})  with the
proviso that unspecified variables should be taken to equal their
most likely values, i.e. those that maximize the entropy $\bar
s(q,\{M_m\})$ not the landscape $\bar \ell(q,\{{\cal
M}_m\})$. For example, in the case of  energy-only landscapes,
$\bar \ell=\bar \ell(\varepsilon)$, the values $q, \{ M_m
\}$ that maximize $\bar s(q,\{M_m\})$ for a given energy
$\varepsilon$ and number of hyperedges $\gamma N\,(\sum_{m=0}^{K}M_m\equiv\gamma)$
 should be computed and then substituted into the expression for
$\bar\ell$ (\ref{ell2}).

\begin{figure}[!ht]
\begin{minipage}[b]{0.45\linewidth}
\centering
\includegraphics[width=2.5in]{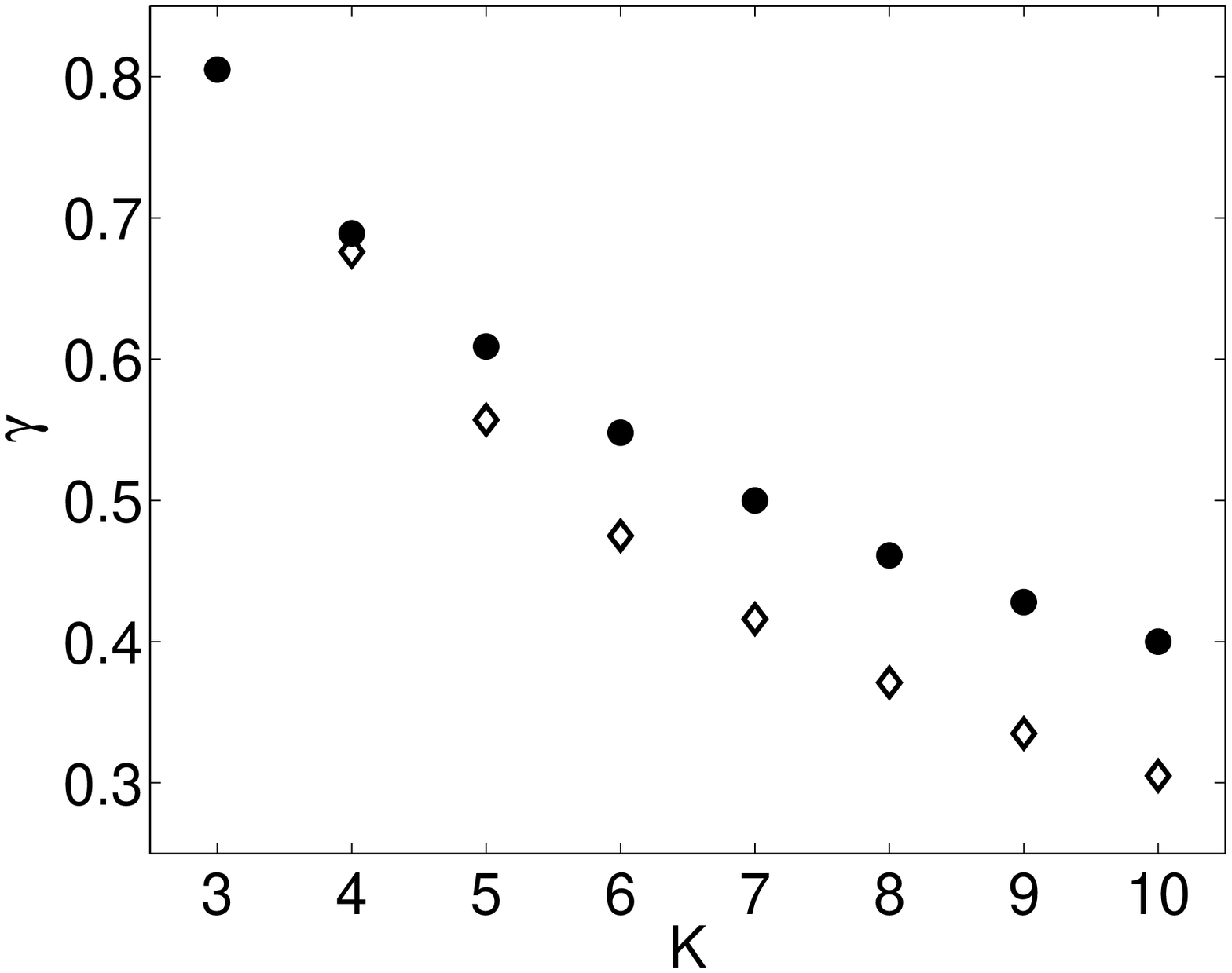}
\caption{\label{fig:1inK} Static $\gamma_c$ (circles) and dynamic
$\gamma_d$ (diamonds) transition for positive 1-in-K SAT for various
values of $K$.}
\end{minipage}
\hspace{0.2in}
\begin{minipage}[b]{0.45\linewidth}
\centering
\includegraphics[width=2.5in]{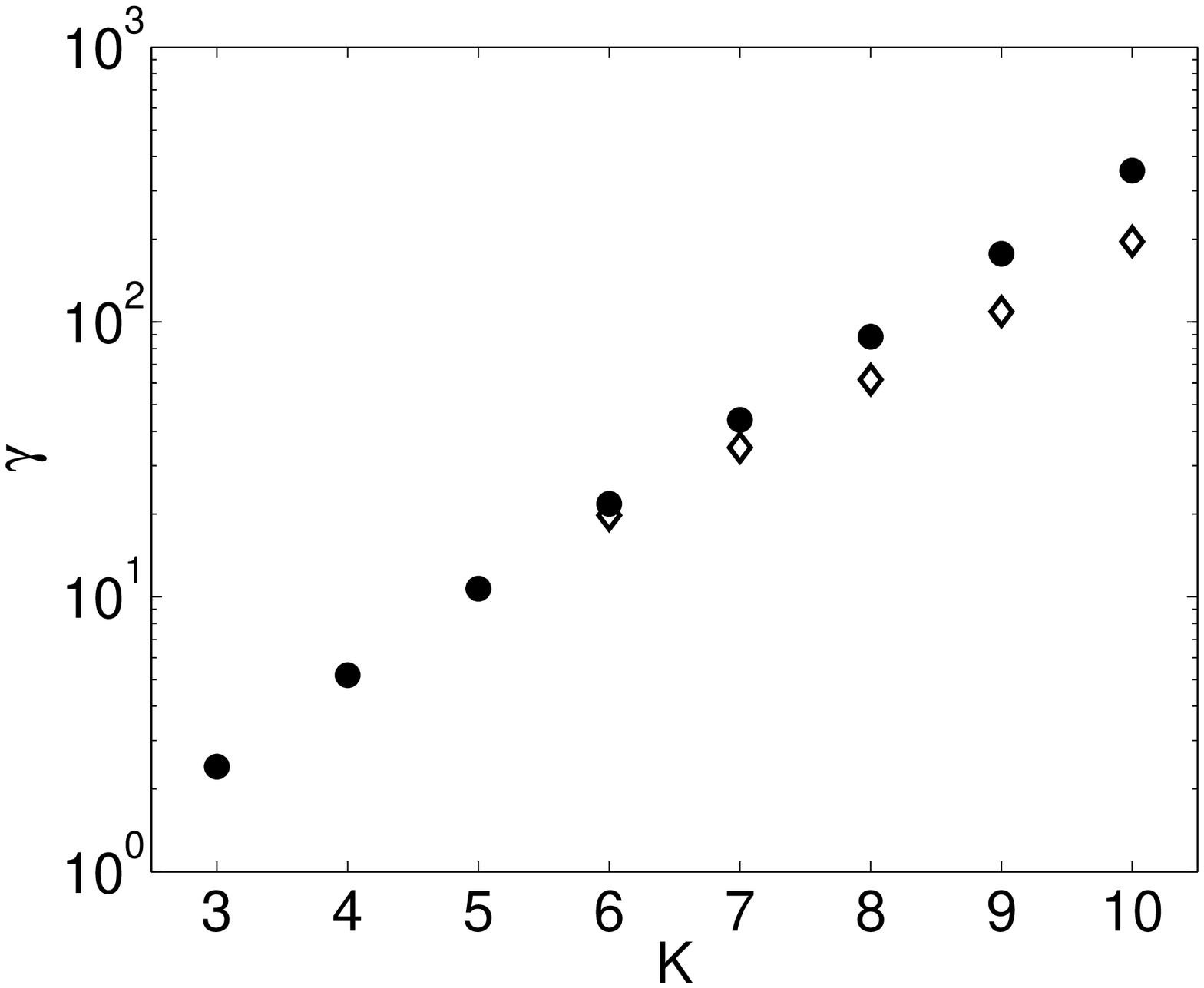}
\caption{\label{fig:KNAE} Static $\gamma_c$ (circles) and dynamic
$\gamma_d$ (diamonds) transition for positive K-NAE SAT for various
values of $K$.}
\end{minipage}
\end{figure}

We compute, within the annealing approximation, the point of
static transition $\gamma_c$ (cf. Fig.\ref{fig:complexity}), where
the entropy  of the macroscopic state with zero energy vanishes,
$s(0)=0$, and the dynamic transition $\gamma_d$; for
connectivities  $\gamma>\gamma_d$ an effective potential
(\ref{FQA}) exhibits a global bifurcation for some
$\Gamma=\Gamma_{\ast}$. The resulting values are given in Table
\ref{tbl:results} (see also  Figs. \ref{fig:1inK} and
\ref{fig:KNAE}). Note that in 1-in-3 SAT and K-NAE-SAT for (K=3,4,5) we
find no dynamical phase transition before the satisfiability threshold (cf. Fig. \ref{fig:complexity}).
\begin{table}[!ht]
\begin{tabular}{|c|c||c|c|c|c|c|c|c|c|}
\hline
& K & 3 & 4 & 5 & 6 & 7 & 8 & 9 & 10 \\
\hline
\raisebox{-0.1in}
{1-in-K} & $\gamma_d$ &   --  & 0.650 & 0.557 & 0.475 & 0.416 & 0.371 & 0.335 & 0.305 \\ [-0.1in]
         & $\gamma_c$ & 0.805 & 0.676 & 0.609 & 0.548 & 0.500 & 0.461 & 0.428 & 0.400 \\
\hline
\raisebox{-0.1in}
{K-NAE} & $\gamma_d$ &  --  &  --  &  --  & 19.8 & 34.9 & 61.7 & 109 & 196 \\ [-0.1in]
        & $\gamma_c$ & 2.41 & 5.19 & 10.7 & 21.8 & 44.0 & 88.4 & 177 & 355 \\
\hline
\end{tabular}
\caption{\label{tbl:results}Annealing bounds for dynamic
($\gamma_d$) and static ($\gamma_c$) transition for positive
1-in-K SAT and positive K-NAE SAT for different values of the
number of variables in a clause $K$. No value (--) indicates the
absence of dynamical transition.}
\end{table}

\begin{figure}[!ht]
\centering
\includegraphics[width=4in]{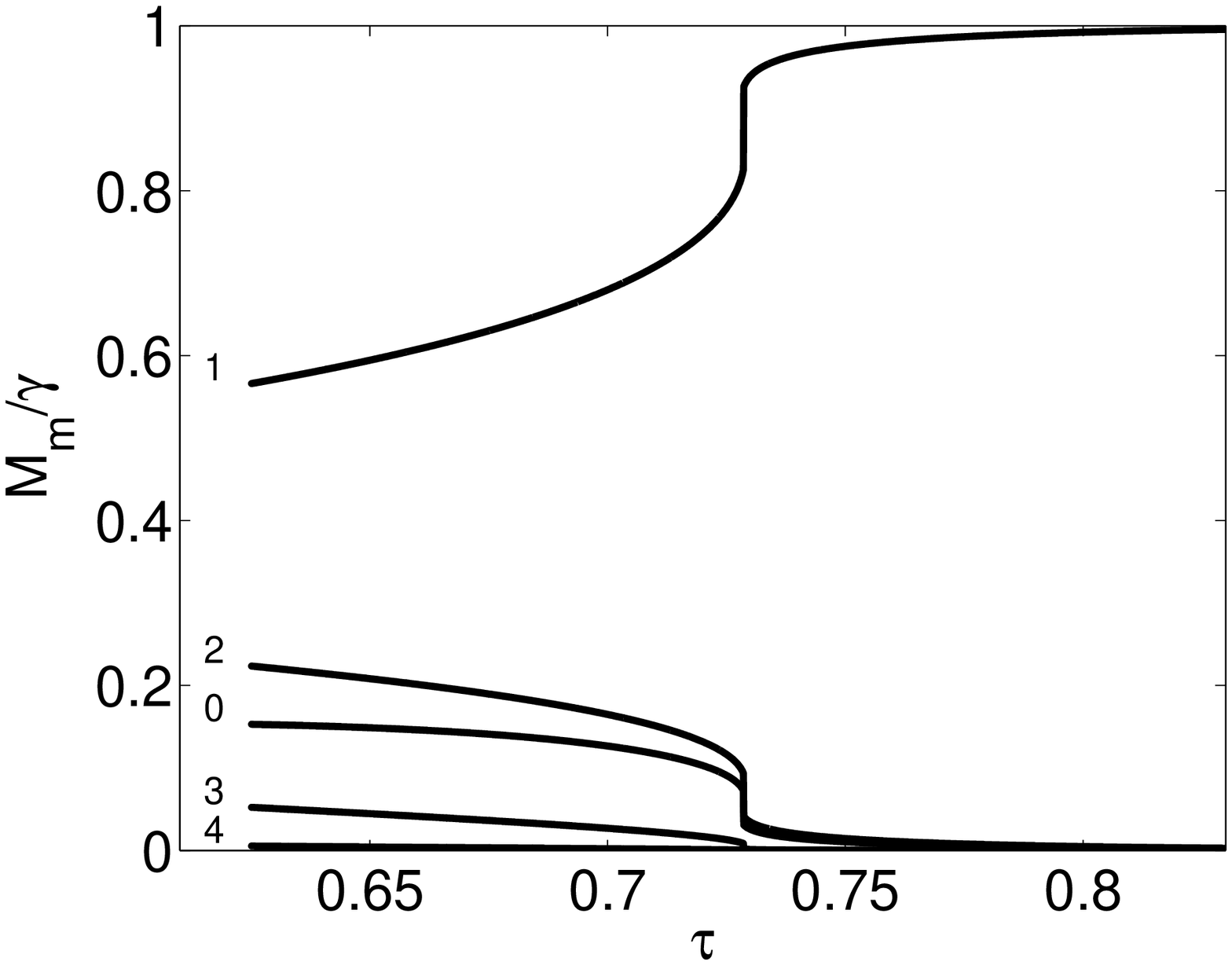}
 \caption{\label{fig:Mms}Plots of the landscape
parameters $M_m=M_{\ast m}$ at the global minimum of
the effective potential,
 {\em vs} $\tau$ for $K=4$ and $\gamma=\gamma_c$ (1-in-K SAT problem). Curves
labeled 0-4 correspond to $M_{\ast 0}/\gamma$ through
$M_{\ast 4}/\gamma$.}
\centering
\includegraphics[width=4in]{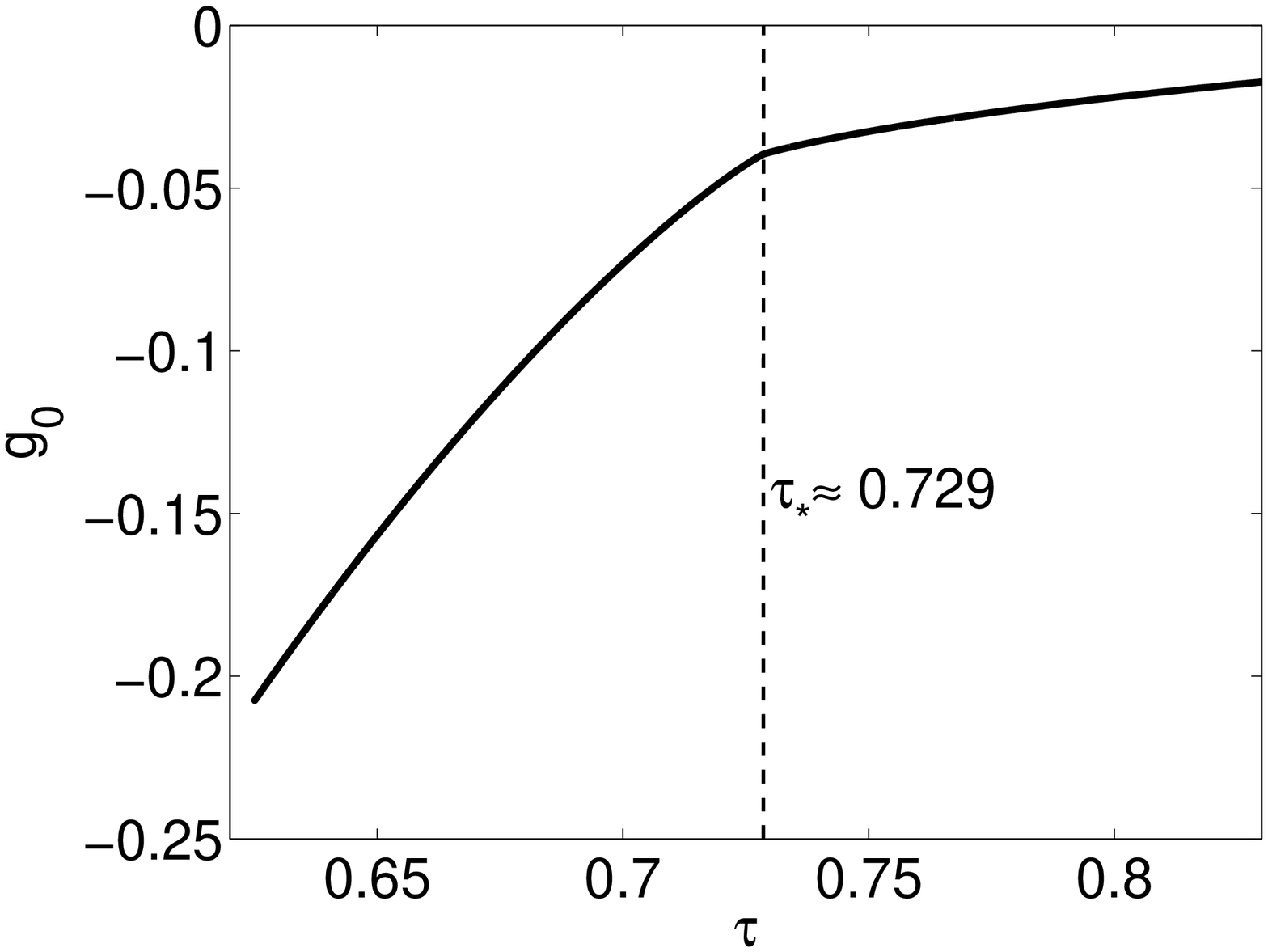}
\caption{\label{fig:F} Scaled energy of adiabatic ground state $g_0$
{\em vs} $\tau$} for K=4 and $\gamma=\gamma_c$ (1-in-K SAT
problem).
\end{figure}

In Fig.~\ref{fig:Mms} we plot  time variations of the landscape
parameters, $M_m=M_{\ast m}$, corresponding to the
global minimum of the effective potential. In Fig. \ref{fig:F} we
plot a time-variation of the scaled ground-state energy $g$ given
by the value of the effective potential at its minimum. Singular
behavior corresponding to the first-order quantum phase transition
at certain $\tau=\tau_*$ ($\Gamma=\Gamma_*$) can be clearly seen
from the figures. Plots in Figs.~\ref{fig:Mms} and \ref{fig:F}
correspond to precisely the static transition $\gamma=\gamma_c$
for the case of $K=4$ in 1-in-K SAT problem.
In the region  $\gamma_d < \gamma < \gamma_c$ there are
exponential (in $N$) number of solutions to Satisfiability problem
but the runtime of the quantum adiabatic algorithm to find any of
them also scales exponentially with $N$. This is a {\em hard}
region for this algorithm.
 We note, that  in the limit of $K
\to \infty$ the annealing approximation becomes exact. Together
with the fact that for large $K$ $\gamma_d$ and $\gamma_c$ seem to
be distinctly different provides evidence that this result
(existence of hard region for quantum adiabatic algorithm) is
robust.

\begin{figure}[b]\hspace{-0.05in}
\includegraphics[width=6.4in]{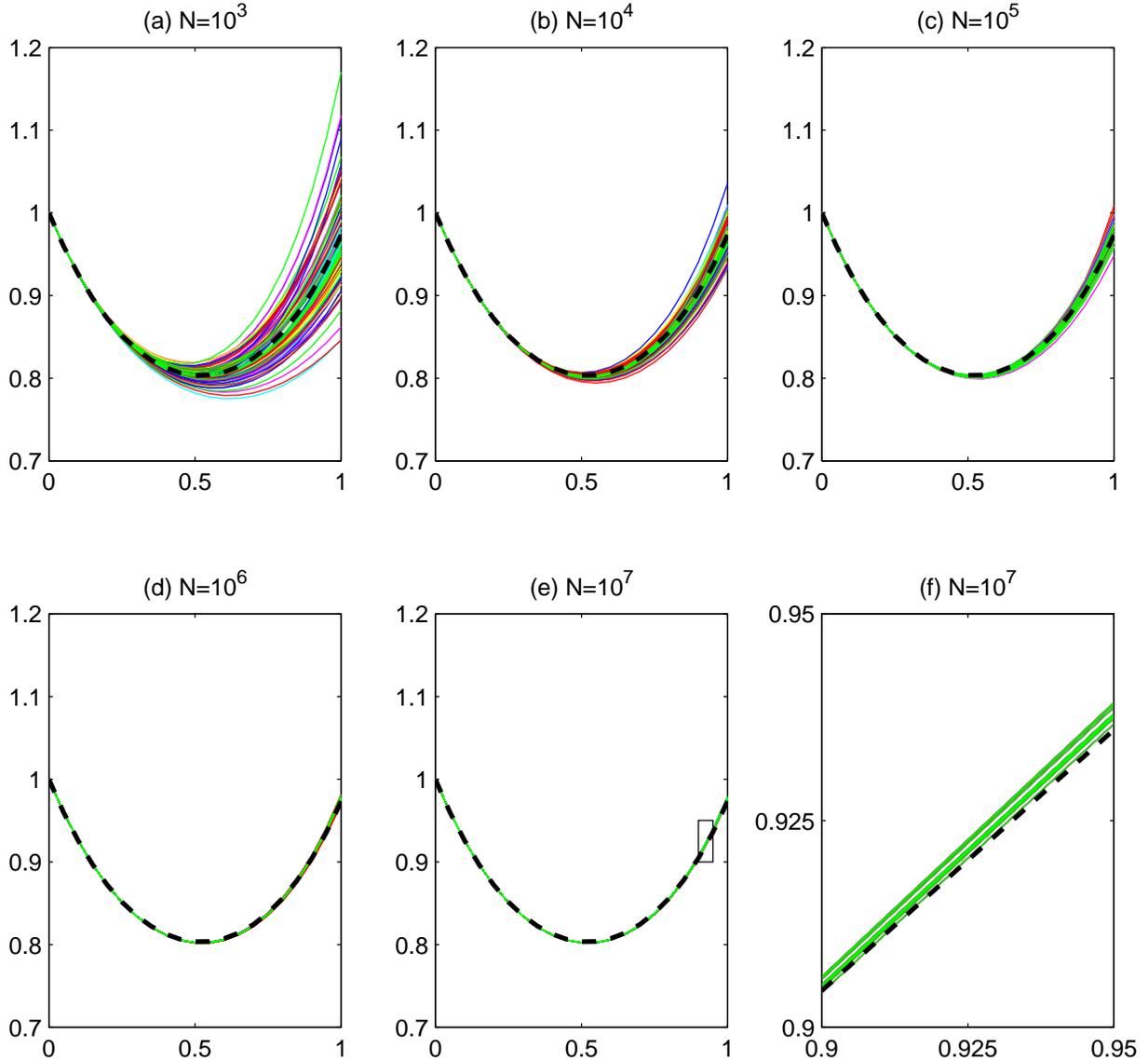}
\caption{\label{fig:simulations} Results of numerical simulations
and their comparison with theory. Depicted are Laplace transforms
of ${\cal M}_1$ for 1-in-3 SAT. Numerical results: curves that have
different colors correspond to  different random problem
instances; curves of same color correspond to different random bit
strings. The dashed black line is a theoretical result based on
the annealing approximation. The insets (a)-(e) depict instances
with $10^3$, $10^4$, $10^5$, $10^6$, and $10^7$ binary variables.
Since the error  is not recognizable we replot in (f) a magnified
section of inset (e). The bit strings were sampled with $q=0.422$,
$M_0=0.048$, $M_1=0.416$, $M_2=0.123$, $M_3=0.013$, corresponding to
$M/N=0.6$.  These values correspond
to the energy $E_\infty/2$ and they are shifted by 10\% from the
most likely values of $q,\{M_m\}$ for this energy (this
shift is $\gg N^{1/2}$). We also note that for 1-in-3 SAT
numerical simulations give static phase transition at
$\gamma_c\approx$ 0.63).}
\end{figure}

\section{\label{sec:Universality}Universality property for transition probabilities}
Here we study the universal features of the transition probability
in (\ref{cond}) for the set of macroscopic variables corresponding
to the (normalized)  total Ising spin $q$ and numbers of clauses
of different types $\{M_m\}$ (\ref{eM}) (the type of a
clause is equal to the number of unit bits  involved in the clause). For
simplicity, we shall focus in this section on the case $K=3$ only.

To clarify the above choice of  macroscopic variables we consider
an auxiliary quantity: a  conditional probability distribution of
the macroscopic variables $(q,\{M_{m}\})$ over the set of
all possible configurations ${\boldsymbol\sigma}$ obtained by
flipping $r$ bits of the configuration
${\boldsymbol\sigma}^{\prime}$.  The first moments of this
distribution corresponding to $M_m$,
\begin{equation}
  \mu_m  =\binom{N}{r}^{-1}\sum_{{\boldsymbol\sigma}}
\delta\left[d({\boldsymbol \sigma}^{\prime}-{\boldsymbol
\sigma}),\,r\right]\,  M_m({\boldsymbol \sigma},{\cal I}),\quad
m=0,\ldots,K,
\end{equation}
\noindent  can be easily computed by counting the number of ways
one can flip $r$ bits in  configuration
${\boldsymbol\sigma^{\prime}}$ to transform a $K$-bit clause of
 $m^{\prime}$ type (i.e., with $m'$ unit bits) into a clause of
the $m$-th type
\begin{equation}
\mu_m=\binom{N}{r}^{-1}\sum_{p,m^{\prime}=0}^{K} M_{m^{\prime}}\,
\binom{m^{\prime}}{p}\binom{K-m^{\prime}}{m-m^{\prime}+p}\binom{N-K}{r-2p-m+m^{\prime}},\label{first}
\end{equation}
\noindent (here we use the convention $\binom{n}{m}\equiv 0$ for
$m<0$ and $m>n$). In the double sum above values of ${\cal
M}_{m'}$ are multiplied by the number of possible ways to flip
three groups of bits: $p$ unit bits in a clause of
$m^{\prime}$-type,  $p+m-m'$ zero bits of this clause, and
$r-2p-m+m'$ bits of the configuration
${\boldsymbol\sigma}^{\prime}$ that do not belong to the clause.
Similarly, one can show that the first moment corresponding to the
variable $q$ equals $q^{\prime}(1-2r/N)$. It is clear that
dependence of the first moments  on the ancestor configuration
${\boldsymbol \sigma^{\prime}}$ is only via the variables
$q^{\prime},\,M_{m}^{\prime}$ for that configuration.

In the limit, $r\gg 1$, the above conditional distribution  has a
Gaussian form with  respect to $q$ and $M_{m}$. Elements of
the covariance matrix
$\Sigma_{m\,q}^{m'\,q'}({\boldsymbol\sigma^{\prime}})={\cal
O}(r)$, and correspondingly, the characteristic width of the
distribution is ${\cal O}(r^{1/2})$. For a configuration
${\boldsymbol\sigma^{\prime}}$ randomly sampled in the box
 $(q,\{M_{m}\})$ the
r.m.s.  deviation of the elements of
$\Sigma_{m\,q}^{m'\,q'}({\boldsymbol\sigma^{\prime}})$ from their
mean values in the box is ${\cal O}(N^{1/2})$. It is clear that in
the limit $r\gg N^{1/2}$ the covariance matrix elements can be
replaced by their mean values for the macroscopic state
$(q,\{M_{m}\})$. Therefore in this limit the conditional
distribution after $r$ spin flips starting from some macroscopic
state depends only on the values of $(q,\{M_{m}\})$ in this
state (universality property).

One can show that for  $r\ll N^{1/2}$  the conditional
distribution after $r$ spin flips can be expressed via the
distribution (\ref{cond}) with  $r=1$, using a standard
convolution rule. Implicit in our derivation of landscapes
is the relation between $r=1$ landscapes and a set of quantities
$\{c_{\sigma,{\bf k}}\}$. Universality of landscapes should be
interpreted as the fact that $\{c_{\sigma,{\bf k}}\}$ are self-averaging.
Had we included only the energy instead of a full set of parameters,
we would not have expected to see such self-averaging in the so-called
replica-symmetry-broken phase. It is possible that inclusion of the
full set of landscape parameters assures universality. We performed a
series of numerical studies to test this hypothesis. In Figure
\ref{fig:simulations} we present the results of numerical simulations
and the comparison with analytic results within the annealing approximation.
One can see that the property of self-averaging holds and that even annealing
approximation provides very good description.

\section{\label{sec:Core}Random graph ensembles with reduced fluctuations}

One can easily point out to a major deficiency of the annealing approximation
-- the fact that it fails to correctly predict the satisfiability transition.
While part of it can be attributed to the fact that real entropy is slightly
different from what is predicted by annealing approximation, major source of
error is the incorrect assumption that the entropy vanishes at the
satisfiability transition. That the entropy does not vanish can be easily seen
by examining the structure of the random graph. At any finite connectivity
(above percolation) it consists of one giant component and a large ${\cal O} ( N )$
number of small ${\cal O} ( 1 )$ components. Each small component contributes ${\cal O} ( 1
)$ contribution to the entropy for a total of ${\cal O} ( N )$ contribution, hence
the entropy is in fact positive at all connectivities, including the
satisfiability threshold.

\subsection{Concept of a {\em core}}

An improvement over annealing bound for satisfiability threshold is possible.
Note that clauses outside of giant component do not affect satisfiability and
hence can be disregarded. Similarly, one can identify {\em irrelevant}
clauses and remove them. Irrelevant clauses are those that can be eliminated
without changing the satisfiability of the entire problem. We shall illustrate
the identification of irrelevant clauses based on local properties for
positive $K$-NAE SAT and positive $1$-in-$K$ SAT.

\subsubsection{Positive $K$-NAE SAT}

\begin{figure}[!ht]
\includegraphics[width=2.5in]{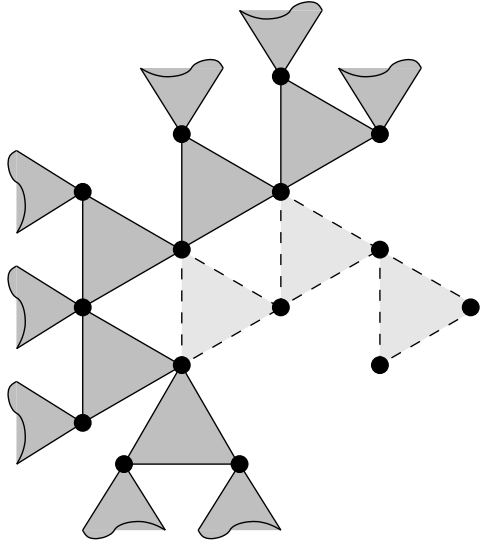}
\caption{\label{fig:trim1}Example of trimming algorithm for 3-NAE SAT. Variables are
represented graphically as vertices and clauses are represented as triangles. Incomplete
triangles represent connections to the remainder of the graph (not shown). Shaded clauses
are removed by the trimming algorithm.}
\end{figure}

For $K$-NAE SAT we can identify variables that do not enter any
clause and remove them without affecting the satisfiability of the
formula. Any variable that appear in {\em exactly one} clause can
also be eliminated together with such clause, since the value of
that variable can be adjusted to satisfy that clause. As one
removes such clause, other variables may become candidates for
removal. One can write an algorithm that iteratively removes
variables that appear in no or exactly one clauses until no such
variables remain (see Fig. \ref{fig:trim1}). In fact using such
algorithm improves running time of classical algorithm. The cost
of this additional preprocessing is negligible: by using special
data structures this algorithm can be made to run in $O ( N \ln N
)$ time.

It is not surprising that the result of running such {\em trimming}
algorithm is a set of clauses and variables with a condition that every
variable appears in at least two clauses. However, the statistical properties
of the remaining {\em core} are not immediately apparent. As a first step,
observe that the resulting core is independent of the order in which
clauses and variables were removed. Hence we can study one specific algorithm.

First, we eliminate all variables of degree $0$ (that appear in no
clauses). In the initial problem all instances with $N$ variables
and $M$ clauses were equally probable. The number of vertices of
degree $0$, $N_0$ does fluctuate, but for {\em every specific}
value of $N_0$ all instances with $N - N_0$ variables, and $M$
clauses, and the additional property that each variable appear in
at least one clause, are equiprobable. This property is referred
to as {\em uniform randomness}. As a next step, identify variables
of degree $1$ (appearing in exactly one clause). Although their
number $N_1$ also fluctuates, for any fixed $N_1$, $N$ and $M$ all
instances with $N$ variables of which $N_1$ variables have degree
$1$ and $M$ clauses are equiprobable.

At each subsequent step of the algorithm we choose {\em at random} a
variable of degree $1$ among $N_1$ candidates, and delete it together with the
clause in which it appears. If the degree of any other variable in that clause
becomes $0$, it is also deleted. Some variables which had degree $2$
previously may become degree $1$ variables. Obviously $M' = M - 1$, but the
values of $N'$ and $N_1'$ cannot be predicted. But since the variable chosen
for deletion was chosen at random (among all degree $1$ variables), one can
show that although the values of $N'$ and $N_1'$ cannot be predicted, for
every fixed set of $\{ N', N_1', M' \}$ all instances are equiprobable.

At the end of the algorithm $N_1 \equiv 0$, hence for fixed $N'$, $M'$, all
instances with $N'$ variables, $M'$ clauses and the condition that each
variable appear in at least two clauses are equiprobable. Since average
changes in $N$, $N_1$ and $M$ at each step were ${\cal O} ( 1 )$ and corresponding
deviations were also ${\cal O} ( 1 )$, after ${\cal O} ( N )$ steps needed for completion of
the algorithm, fluctuations in resulting $N'$ and $M'$ are only ${\cal O} ( \sqrt{N} )$
by central limit theorem and can be neglected in our analysis.

There are several methods to derive the average resulting values of $N'$,
$M'$. The most straightforward is to study the evolution of $N$, $N_1$ and $M$
by solving differential equations for their average values. This is somewhat
tedious and less general. We instead resort to a different method of
self-consistency equations. For every instance one can identify a set
${\cal C}$ of variables that form a {\em core}, i.e. those that will not
be delete by the trimming algorithm. We now extend that set to ${\cal C}'$
using the following rules:
\begin{enumerate}
  \item If a variable belongs to ${\cal C}$, it also belong to
  ${\cal C}'$.

  \item If $K - 1$ variables in some clause belong to ${\cal C}'$, then the
  remaining variable must also belong to ${\cal C}'$.
\end{enumerate}
The minimal set ${\cal C}'$ is obtained from ${\cal C}$ by iteratively
adding variables according to these rules. Let the number of variables in set
${\cal C}'$ be equal to $qN$ with $q < 1$. Fluctuations in $q$ are $O ( 1 /
\sqrt{N} )$ and, consequently, are neglected.

Now introduce $( N + 1 )$-st variable and compare systems with $N$ variables
and $N + 1$ variables. Together with $( N + 1 )$-st variable introduce $\Delta
M$ clauses each involving that variable and $( K - 1 )$ random variables. The
number $\Delta M$ is Poisson random variable with parameter $K \gamma$. We can
compute the probability that the new variable belongs to ${\cal C}'$ of the
enlarged instance. The new variable belongs to ${\cal C}'$ if for at least
one of $\Delta M$ clauses, $K - 1$ variables other than the new variables are
not in ${\cal C}'$ at the same time. Since these clauses are connected to
random vertices, that probability can be expressed via $q$ alone and, owing to
the fact that $\Delta M$ is Poisson, equals $1 - \exp ( - K \gamma q^{K - 1}
)$. Self-consistency requires that this probability be equal to $q$:
\begin{equation} 1 - q = e^{- K \gamma q^{K - 1}} . \end{equation}
For practical purposes we must seek the largest solution to this equation.

Note that the new variable belongs to the core (set ${\cal C}$) if not for
one, but for {\em at least two} of $\Delta M$ clauses, $K - 1$ variables
other than the new one are in ${\cal C}'$. The probability of that
\begin{equation} p = 1 - e^{- K \gamma q^{K - 1}} - K \gamma q^{K - 1} e^{- K \gamma q^{K -
   1}} \equiv q - K \gamma q^{K - 1} ( 1 - q ) \end{equation}
is equal to the average value of $N' / N$, and the average connectivity of the
new vertex (under the condition that it is in ${\cal C}$) enables to
compute $M' / N'$:
\begin{equation} M' / N' = \frac{1}{K}  \frac{K \gamma q^{K - 1} \left( 1 - e^{- K \gamma
   q^{K - 1}} \right)}{1 - ( 1 + K \gamma q^{K - 1} ) e^{- K \gamma q^{K -
   1}}} \equiv \frac{\gamma q^K}{p} \end{equation}
(better expressed as $M'/N=\frac{1}{K} K \gamma q^{K-1} \left(
1 - e^{-K \gamma q^{K-1}} \right) \equiv \gamma q^K$).

We must mention that the core for $K$-NAE SAT is in fact identical to that
for $K$-XOR SAT. The latter has been analyzed by a different method \cite{Mezard:02}.

\subsubsection{Positive $1$-in-$K$ SAT}

The situation is slightly more interesting in case of positive $1$-in-$K$ SAT.
Setting a variable that appears in only one clause does not guarantee that the
clause will be satisfied, hence the clause cannot be eliminated. Although it
cannot be eliminated, it can still be simplified by eliminating the variable.
The remaining instance will contain not only $K$-clauses but also one $(K-1)$-clause.
The allowed combinations of variables for $(K-1)$-clauses
should be those that have sum of variables equal to {\em either} $0$
{\em or} $1$. In that case the value of the deleted variable can always be
adjusted so that the sum of variables in $K$-clause is exactly $1$.

\begin{figure}[!ht]
\includegraphics[width=6in]{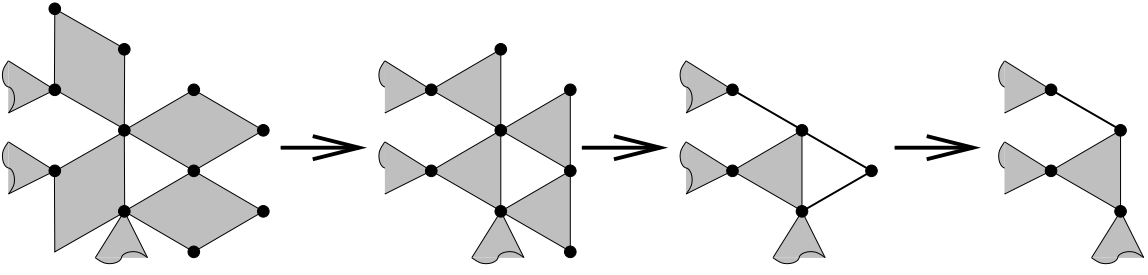}
\caption{\label{fig:trim2}Example of trimming algorithm for 1-in-4 SAT. Variables are represented
graphically as vertices and 4-, 3- and 2-clauses are represented as rhombi, triangles
and edges correspondingly. Incomplete polygons represent connections to the remainder of the graph.
The figure depicts evolution of part of the graph under trimming algorithm.}
\end{figure}

Similarly, at some point $(K-1)$-clauses can be converted to $(K-2)$-clauses,
$(K-3)$-clauses, down to $2$-clauses. In each case the allowed
combinations shall have sum of spins either $0$ or $1$. Finally, if one can
identify a variable that appears in $2$-clauses exclusively, this variable can
be eliminated together with corresponding $2$-clauses, since setting it to $0$
shall satisfy all these clauses. As an illustration, see Fig. \ref{fig:trim2}
which depicts $K=4$.

Once again, one can write trimming algorithm that eliminates such variables
and clauses. As before, one can show that all remaining instances must have
the property that each variable appear in at least two clauses of any
length and at least one clause of length greater than two; and that for fixed
number of vertices $N'$, number of $2$-clauses $M_2'$, ..., $(K-1)$-clauses
$M_{K-1}'$, $K$-clauses $M_K'$, all such instances are equiprobable.

Without going into details (available in our upcoming paper \cite{upcoming}), we should
mention that equations now involve two variables $q$ and $q'$, owing to
separate treatment afforded to $2$-clauses and clauses of length greater than
$2$. The self-consistency equations are:
\begin{eqnarray}
  1-q' & = & e^{- K \gamma \left[1-(1-q)^{K-1}-(K-1)q(1-q)^{K-2} \right]},\\
  1-q & = & (1-q') e^{- K (K-1) \gamma q'(1-q)^{K-2}},
\end{eqnarray}
with implicit understanding that the largest solutions $q$, $q'$ are sought.

The expected values of $N'$ as well as $M_k'$ are given by
\begin{eqnarray}
  N'/N & = & q' - K \gamma \left[ 1-(1-q)^{K-1}-(K-1)q(1-q)^{K-2} \right] (1-q),\\
  M_2'/N & = & \binom{K}{2} \gamma q'^2 (1-q)^{K-2},\\
  M_k'/N & = & \binom{K}{k} \gamma q^k (1-q)^{K-k} \quad {\rm for } k \geqslant 3.
\end{eqnarray}
Note that results for the core for positive $K$-NAE SAT and positive
$1$-in-$K$ SAT allow for a compact formulation through introduction of
{\em generating function} of allowed variable degrees.

For the case of $K$-NAE SAT it is $G ( \mu ) = \sum_{d \geqslant 2} \mu^d / d!
= e^{\mu} - 1 - \mu$, since degrees of all variables are at least $2$. In
terms of this $G ( \mu )$, equations can be written as
\begin{eqnarray}
  N' / N & = & e^{- \mu} G ( \mu )\\
  M' / N & = & \frac{1}{K} \mu e^{- \mu} G' ( \mu )
\end{eqnarray}
with $\mu = K \gamma q^{K - 1}$, and the equation on $q$ is
\begin{equation} q = e^{- \mu} G' ( \mu ) . \end{equation}
For the case of $1$-in-$K$ SAT, $G$ is a function of $( K - 1 )$ variables
reflecting the fact that the degree of each vertex is a vector $( d_2, \ldots
., d_K )$ where $d_k$ counts the number of appearances of said variable in
$k$-clauses. The generating function for the remaining core is
\begin{equation} G ( \mu_2, \mu_3, \ldots, \mu_K ) = e^{\sum_{i = 2}^K \mu_i} - e^{\mu_2} -
   \sum_{i = 3}^K \mu_i . \end{equation}
There appears a set $\{ q_2, \ldots, q_K \}$ such that $q_k = ( \partial /
\partial \mu_k ) G ( \{ \mu_i \} )$. Obviously, $q_3 = q_4 = \ldots = q_K$.
Setting
\begin{eqnarray}
  \mu_2 & = & 2 \binom{K}{2} \gamma q'(1-q)^{K-2},\\
  \mu_k & = & k \binom{K}{k} \gamma q^{k-1}(1-q)^{K-k} \quad {\rm for} k \geqslant 3,
\end{eqnarray}
$N'$ and $M_k'$ can be compactly written as
\begin{eqnarray}
  N'/N & = & e^{- \sum_{i = 2}^K \mu_i} G ( \{ \mu_i \} ),\\
  M_k'/N & = & \frac{1}{k} \mu_k e^{- \sum_{i = 2}^K \mu_i}
               \frac{\partial}{\partial \mu_k} G ( \{ \mu_i \} ) .
\end{eqnarray}

\subsection{Improved annealing bound}

Once irrelevant clauses have been removed we expect the entropy at the
satisfiability transition to be much closer to zero. We have already made
predictions for the number of remaining variables and clauses after trimming
algorithm starting from a random graph of connectivity $\gamma$. We can
compute the entropy of the remaining core. Using the critical connectivity at
which the annealed entropy of the core alone becomes $0$ as an estimate of
satisfiability transition is an improvement over traditional annealing
approximation. We are motivated by two contributing factors: (i) the rigorous
mathematical proof that the disorder is relevant for satisfiability transition
relies on presence of irrelevant clauses, hence their removal can sometimes
make disorder irrelevant, and (ii) applied to $K$-XOR SAT, this improved bound
becomes {\em exact} \cite{Mezard:02}.

Doing annealing approximation on a core is hampered by the fact that clauses
can no longer be assumed uncorrelated. However, the technique we introduced in
this paper is well-suited for this example. The annealed entropy equals
\begin{equation}
   s_{\rm ann} = \frac{1}{N'} \ln {\cal N}( s, J ) - \frac{1}{N'} \ln {\cal N}( J ),
\end{equation}
where ${\cal N}( J )$ counts the number of possible disorders and
${\cal N}( s, J )$ is the total number of allowed spin-disorder
configurations. We now separately consider $K$-NAE SAT and $1$-in-$K$ SAT.

\subsubsection{Positive $K$-NAE SAT}

Consider the number of possible disorder realizations with $N'$ variables,
$M'$ clauses with the constraint that each variable appear in at least two
clauses. Each variable can be characterized by the vector ${\bf k}$ with
$K$ components, each component $k_p$ counting the number of clauses where said
variable appears in $p$-th position. Treating disorders that differ by
permuting different clauses or permuting variables within a clause as
different and fixing the fractions of vertices with particular realization of
${\bf k}$ to $c_{\bf k}$, the normalized logarithm of the count of
possible disorders is
\begin{equation}
   \frac{1}{N'} \ln {\cal N}( J ) \equiv s_0' [ N', M' ; \{
   c_{\bf k} \} ] = - \sum_{\bf k} c_{\bf k} \ln \left[
   c_{\bf k}  \prod_p k_p ! \right] + K \frac{M'}{N'} \ln M' .
\end{equation}
We must maximize this expression with respect to $c_{\bf k}$ subject to
constraints
\begin{equation} \sum_{\bf k} k_p c_{\bf k} = M' / N' \end{equation}
and that $c_{\bf k} = 0$ for $\sum_p k_p < 2$.

Introducing Lagrange multipliers $\mu_p$, the expression for
$c_{\bf k}$ becomes
\begin{equation}
   c_{\bf k} = \frac{1}{G ( \sum_p \mu_p )}  \prod_p
   \frac{\mu_p^{k_p}}{k_p !} .
\end{equation}
Echoing the constraint that $\sum_p k_p \geqslant 2$, the
normalization factor is $G ( \sum_p \mu_p )$ where $G ( x ) = e^x
- 1 - x$ -- the familiar generating function of the core. Using
dual transformation we can rewrite the entropy as
\begin{equation}
   s_0' [ N', M' ] = \min_{\mu_p} \left\{ \sum_p ( M' / N' ) \ln \frac{M' /
   N'}{\mu_p} + \ln G \left( \sum_p \mu_p \right) \right\} + \frac{K M'}{N'}
   \ln N' .
\end{equation}
Comparing this expression with Eq.() we observe that the minimum is achieved
for $\mu_p = \frac{1}{K} \mu \equiv \gamma q^{K - 1}$, $\gamma$ being the
connectivity of untrimmed random graph.

Now we proceed to count all possible combinations of spin assignments and
disorder realizations ${\cal N}( s, J )$. Corresponding entropy is in
general a function of $\{ {\cal M}_{\boldsymbol \alpha}' \}$ where
${\cal M}_{\boldsymbol \alpha}'$ is the number of clauses with variables having values
described by a vector $\{ \alpha_p \} \equiv {\boldsymbol \alpha}$, divided by $N'$ so that
$\sum_{\boldsymbol \alpha}{\cal M}{\boldsymbol \alpha}=M'/N'$. In complete
analogy with results of section \ref{sec:entropy} we obtain
\begin{eqnarray}
   s_1' [ N', M' ; \{ {\cal M}_{\boldsymbol \alpha}' \} ; \{ c_{\sigma,{\bf k}}
   \} ] & = & - \sum_{\sigma,{\bf k}} c_{\sigma,{\bf k}} \ln \left[
   c_{\sigma,{\bf k}}  \prod_{p,{\bf \alpha}}
   k_{\boldsymbol \alpha}^p ! \right] + (K-1) \sum_{\boldsymbol \alpha}
   {\cal M}_{\boldsymbol \alpha}' \ln {\cal M}_{\boldsymbol \alpha}' +
   \frac{M'}{N'} \ln \frac{M'}{N'} \nonumber \\ && + \frac{K M'}{N'}\ln N'.
\end{eqnarray}
Optimizing this with respect to $c_{\sigma,{\bf k}}$ subject to familiar
constraints
\begin{equation}
   \sum_{\sigma,{\bf k}} k_{\boldsymbol \alpha}^p c_{\sigma,{\bf k}}
   = {\cal M}_{\boldsymbol \alpha}
\end{equation}
and the constraint that $c_{\sigma,{\bf k}} = 0$ if
$\sum_{p,{\boldsymbol \alpha}} k_{\boldsymbol \alpha}^p < 2$, the expression can
be equivalently rewritten as
\begin{eqnarray}
   s_1' [ N', M' ; \{ {\cal M}_{\boldsymbol \alpha}' \} ] & = &
   \min_{\mu_{\boldsymbol \alpha}^p} \left\{ \sum_{p,{\boldsymbol \alpha}}
   {\cal M}_{\boldsymbol \alpha}' \ln
   \frac{{\cal M}_{\boldsymbol \alpha}'}{\mu_{\boldsymbol \alpha}^p} + \ln Z [ \{
   \mu_{\boldsymbol \alpha}^p \} ] \right\} + \frac{M'}{N'} \ln \frac{M'}{N'} -
   \sum_{\boldsymbol \alpha} {\cal M}_{\boldsymbol \alpha}' \ln
   {\cal M}_{\boldsymbol \alpha}' \nonumber \\ && + \frac{K M'}{N'} \ln N',
\end{eqnarray}
with $Z [ \{ \mu_{\boldsymbol \alpha}^p \} ]$ given instead by
\begin{equation}
   Z [ \{ \mu_{\boldsymbol \alpha}^p \} ] = G \left( \sum_{p,{\boldsymbol \alpha}}
   \delta [ \alpha_p ; 1 ] \mu_{\boldsymbol \alpha}^p \right) + G \left(
   \sum_{p,{\boldsymbol \alpha}} \delta [ \alpha_p ; - 1 ]
   \mu_{\boldsymbol \alpha}^p \right) .
\end{equation}
Labeling arguments by $\mu_+$ and $\mu_-$ respectively allows us to simplify
this to
\begin{eqnarray}
   s_1' [ \{ {\cal M}_{\boldsymbol \alpha}' \} ] & = & \min_{\mu_{\pm}} \left\{ {\cal M}_+' \ln
   \frac{{\cal M}_+'}{\mu_+} + {\cal M}_-' \ln \frac{{\cal M}_-'}{\mu_-} + \ln [ G (
   \mu_+ ) + G ( \mu_- ) ] \right\} \nonumber \\
   && + \frac{M'}{N'} \ln \frac{M'}{N'} - \sum_{\boldsymbol \alpha}
   {\cal M}_{\boldsymbol \alpha}' \ln {\cal M}_{\boldsymbol \alpha}' +
   \frac{KM'}{N'} \ln N' .
\end{eqnarray}
with ${\cal M}_+' = \sum_{{\boldsymbol \alpha}} \left( \sum_p \delta [ \alpha_p ; 1 ]
\right) {\cal M}_{{\boldsymbol \alpha}}'$ and ${\cal M}_-' = \sum_{{\boldsymbol \alpha}} \left(
\sum_p \delta [ \alpha_p ; - 1 ] \right) {\cal M}_{{\boldsymbol \alpha}}'$.

The correct annealed entropy is given by the difference of these expressions
\begin{equation}
   s_{{\rm ann}}' [ N', M' ; \{ {\cal M}_{\boldsymbol \alpha}' \} ] = s_1' [ N', M' ;
   \{ {\cal M}_{{\boldsymbol \alpha}}' \} ] - s_0' [ N', M' ]
\end{equation}
This enables us to find the improved bound on satisfiability threshold (where
$H_{{\rm ann}} = 0$) or to compute the most likely values of $\{
{\cal M}_{{\boldsymbol \alpha}}' \}$ for particular energy $E$.

Note that for positive $K$-NAE SAT corrections to static threshold due to this
improved approximation are minute, since the transition happens at large connectivities,
where simple annealing approximation adequately describes the transition.

\subsubsection{Positive $1$-in-$K$ SAT}

For $1$-in-$K$ SAT the derivation is quite similar. An important addition is
that the clauses can have any length from $2$ to $K$, hence index
${\bf k}$ should reflect that fact. Since calculations are similar in
spirit, we shall only provide the results
\begin{equation}
   s_0' [ N', \{ M_k' \} ] = \min_{\mu_k} \left\{ k \frac{M_k'}{N'} \ln
   \frac{k M_k' / N'}{\mu_k} + \ln G ( \{ \mu_k \} ) \right\} + \sum_{k = 2}^K
   \frac{k M_k'}{N'} \ln N',
\end{equation}
with $G ( \{ \mu_k \} ) = \exp \left( \sum_{k = 2}^K \mu_k \right) - e^{\mu_2}
- \sum_{k = 3}^K \mu_k$, and
\begin{eqnarray}
   s_1' & = & \min_{\mu_{k, \pm}} \left\{
   \sum_{k = 2}^K \left( {\cal M}_{k, +}' \ln \frac{{\cal M}_{k, +}'
   }{\mu_{k, +}} + {\cal M}_{k, -}' \ln \frac{{\cal M}_{k, -}'}{\mu_{k,
   -}} \right) + \ln [ G ( \{ \mu_{k, +} \} ) + G ( \{ \mu_{k, -} \} ) ]
   \right\} \nonumber \\
   &&+ \sum_{k = 2}^K \left[ {\cal M}_k' \ln {\cal M}_k' -
   \sum_{{\boldsymbol \alpha}} {\cal M}_{k,{\boldsymbol \alpha}}' \ln
   {\cal M}_{k,{\boldsymbol \alpha}}' \right] + \sum_{k = 2}^K
   \frac{k M_k'}{N'} \ln N',
\end{eqnarray}
and the complete expression for the annealed entropy is
\begin{equation}
   s_{{\rm ann}}' [ N', \{ {\cal M}_k' \} ; \{ {\cal M}_{k,{\boldsymbol \alpha}}' \} ] = s_1'
   [ \{ {\cal M}_{k,{\boldsymbol \alpha}}' \} ] - s_0' [ N', \{ M_k' \} ] .
\end{equation}
The numerical predictions for the point where the entropy becomes zero are provided in
Table \ref{tbl:core}.

\section{\label{sec:CoreQAA}QAA algorithm on a core: extended landscapes}

Once we have reexpressed the entropy on a reduced graph, it is only natural to
implement the quantum adiabatic evolution directly on the reduced graph.
Correspondingly, we shall need to recompute landscapes for the reduced graph.

\subsection{Positive $K$-NAE SAT}

This case entails least difficulty, since we can use same landscape parameters
$\{ {\cal M}_{\boldsymbol \alpha}' \}$, the difference being the exclusion of total
spin from the list of parameters. The central quantity--$\ell' ( \{
{\cal M}_{{\boldsymbol \alpha}}' \} )$--is still expressed in a similar form:
\begin{equation}
   \ell' ( \{ {\cal M}_{\boldsymbol \alpha}' \} ) = \sum_{\sigma,{\bf k}}
   c_{\sigma,{\bf k}} \exp \left[ \frac{1}{2} \sum_{p,{\boldsymbol \alpha}}
   k_{\boldsymbol \alpha}^p \left( \frac{\partial s_{\rm ann}'}{\partial
   {\cal M}_{\boldsymbol \alpha}'} - \frac{\partial s_{\rm ann}'}{\partial
   {\cal M}_{\bar{\boldsymbol \alpha} ( p,{\boldsymbol \alpha})}'} \right) \right] .
\end{equation}
Notably, the sum over ${\bf k}$ is now restricted to $|{\bf k}|
\geqslant 2$. The derivatives of the entropy are still
\begin{equation}
   \frac{\partial s_{\rm ann}'}{\partial {\cal M}_{\boldsymbol \alpha}'} = \sum_p
   \ln \frac{{\cal M}_{\boldsymbol \alpha}'}{\mu_{\boldsymbol \alpha}^p} - \ln
   {\cal M}_{\boldsymbol \alpha}' .
\end{equation}
For ${\boldsymbol \alpha}$, ${\boldsymbol \alpha}'$ differing in exactly one
position, we shall have
\begin{equation}
   \frac{1}{2} \left( \frac{\partial s_{\rm ann}'}{\partial
   {\cal M}_{\boldsymbol \alpha}'} - \frac{\partial s_{\rm ann}'}{\partial
   {\cal M}_{{\boldsymbol \alpha}'}'} \right) = \frac{1}{2} \ln
   \frac{\mu_{{\boldsymbol \alpha}'}^p}{\mu_{\boldsymbol \alpha}^p} .
\end{equation}
Next, using $c_{\sigma,{\bf k}} = \frac{1}{Z} ( \mu_{\boldsymbol \alpha}^p
)^{k_{{\boldsymbol \alpha}}^p} / k_{{\boldsymbol \alpha}}^p !$, for $|{\bf k}|
\geqslant 2$ and Eq. (\ref{Laplace}), we are able to write, in terms of
generating function $G(\mu)$:
\begin{equation}
   \ell' ( \{ {\cal M}_{\boldsymbol \alpha}' \} ) = \frac{2}{Z} G \left(
   \sum_{<{\boldsymbol \alpha},{\boldsymbol \alpha}' >}
   \sqrt{\mu_{\boldsymbol \alpha}^p \mu_{{\boldsymbol \alpha}'}^p} \right)
\end{equation}
Substituting ${\cal M}_{\boldsymbol \alpha}' / \mu_{\boldsymbol \alpha}^p =
{\cal M}_{\alpha_p}' / \mu_{\alpha_p}$, we can rewrite this as
\begin{equation}
   \ell' ( \{ {\cal M}_{\boldsymbol \alpha}' \} ) = \frac{2}{G ( \mu_+ ) + G ( \mu_- )} G
   \left( \sqrt{\frac{\mu_+ \mu_-}{{\cal M}_+' {\cal M}_-'}}
   \sum_{<{\boldsymbol \alpha},{\boldsymbol \alpha}' >} \sqrt{{\cal M}_{\boldsymbol \alpha}'
   {\cal M}_{{\boldsymbol \alpha}'}'} \right) .
\end{equation}
Note that for the landscapes on the original graph without spin variable, this
expression is valid with $G ( \mu )$ redefined to be $G ( \mu ) = e^{\mu}$.

\subsection{Positive $1$-in-$K$ SAT}

The notion of landscape parameters has to be generalized, since clauses of any
length can appear. Correspondingly, we choose a set $\{
{\cal M}_{k,{\boldsymbol \alpha}}' \}$ to serve as landscape parameters. In contrast to
$K$-NAE SAT, we have a total of $\sum_{k = 2}^K ( k + 1 ) = ( K - 1 ) ( K - 2
) / 2$ parameters (with symmetries of ${\cal M}_{k,{\boldsymbol \alpha}}'$ taken into
account).

The derivation of landscapes can be generalized to include several types of
clauses. The final answer is itself a generalization of Eq. (\ref{ell1})
\begin{equation}
   \ell' ( \{ {\cal M}_{k,{\boldsymbol \alpha}}' \} ) = \frac{2}{G ( \{ \mu_{k, +} \} ) + G
   ( \{ \mu_{k, -} \} )} G \left( \left\{ \sqrt{\frac{\mu_{k, +} \mu_{k,
   -}}{{\cal M}_{k, +} {\cal M}_{k, -}}} \sum_{<{\boldsymbol \alpha},{\boldsymbol \alpha}' >}
   \sqrt{{\cal M}_{k,{\boldsymbol \alpha}}' {\cal M}_{k,{\boldsymbol \alpha}'}'} \right\} \right).
\end{equation}
As before $G ( \{ \mu_k \} ) = \exp \left( \sum_{i = 2}^K \mu_i \right) -
e^{\mu_2} - \sum_{i = 3}^K \mu_i$.

We can exploit the symmetry of ${\cal M}_{k,{\boldsymbol \alpha}}'$ to write
\begin{equation}
   \sum_{<{\boldsymbol \alpha},{\boldsymbol \alpha}' >}
   \sqrt{{\cal M}_{k,{\boldsymbol \alpha}}' {\cal M}_{k,{\boldsymbol \alpha}'}'} = \sum_{m = 0}^{k
   - 1} \sqrt{( m + 1 ) ( k - m ) M_{k, m}' M_{k, m + 1}'} .
\end{equation}
Same is possible for $K$-NAE SAT; in fact it is precisely this form that is
used in numerical calculations.

\subsection{Numerical Results}

Here we provide the numerical results for the satisfiability transition as
determined by maximizing the entropy for energy $E = 0$ and solving
$s_{\rm ann}' = 0$. And we also list the location of dynamical transition,
indicated by global bifurcation in $f' = \varepsilon' - \Gamma \ell'$. All
results are expressed in terms of connectivity of the original random graph
for easy comparison with Table \ref{tbl:results}.

\begin{figure}[!ht]
\includegraphics[width=3.5in]{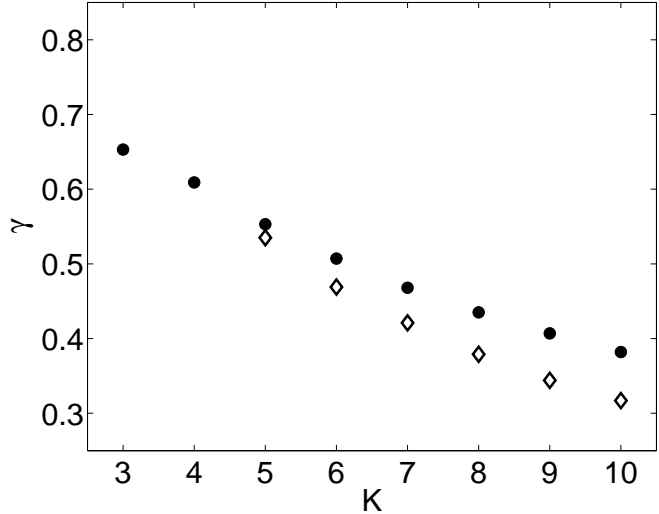}
\caption{\label{fig:core} Static (circles) and dynamic (diamonds) transition for various values
of $K$ for positive $1$-in-$K$ SAT.}
\end{figure}

We observed that this refinement of our analysis leaves $\gamma_d$ and $\gamma_c$
of $K$-NAE SAT essentially unchanged. This is the manifestation of the fact that
if either $K$ or $\gamma$ is large, the core is not much different from the original
graph (i.e. that $q\approx 1$). In contrast, the difference for $1$-in-$K$ SAT is
quite perceptible. For consistency, we compare the location of dynamic phase transition
computed on the core to that computed on the original graph using only
${M_m}$ as landscape parameters (omitting total spin, as it is not among parameters
for computations performed on a core). We also include new better bounds on static
transition. All results are summarized in Table \ref{tbl:core} and Fig. \ref{fig:core}.

\begin{table}[!ht]
\begin{tabular}{|c|c||c|c|c|c|c|c|c|c|}
\hline
& K & 3 & 4 & 5 & 6 & 7 & 8 & 9 & 10 \\
\hline
\raisebox{-0.1in}
{Improved} & $\gamma_d'$ &   --  &   --  & 0.535 & 0.469 & 0.421 & 0.379 & 0.344 & 0.317 \\ [-0.1in]
           & $\gamma_c'$ & 0.653 & 0.609 & 0.553 & 0.507 & 0.468 & 0.435 & 0.407 & 0.382 \\
\hline
\raisebox{-0.1in}
{Old} & $\gamma_d$ &  --  &  0.671 & 0.552 & 0.471 & 0.413 & 0.368 & 0.333 & 0.304 \\ [-0.1in]
      & $\gamma_c$ & 0.805 & 0.676 & 0.609 & 0.548 & 0.500 & 0.461 & 0.428 & 0.400 \\
\hline
\end{tabular}
\caption{\label{tbl:core}
Dynamic and static transition for positive $1$-in-$K$ SAT with improved annealing approximation
and with old method using $\{M_m\}$ as landscape parameters. Prediction of $\gamma_c'$ for $K=3$ compares favorably with
result of simulations of $\gamma_c' \approx 0.63$. No value (--) indicates the ebasence of dynamical transition.}
\end{table}

\section{\label{sec:Conclusion}Conclusion}

We have formulated an ansatz of landscapes and studied the
complexity of the quantum adiabatic algorithm within the annealing
approximation and found the existence of a dynamic transition and
a hard(exponential) region above that dynamic transition. However,
a similar analysis of simulated annealing did not reveal any phase
transitions. We explain this as follows. The annealing
approximation should fail for sufficiently small energies. It is
commonly known that simulated annealing can find suboptimal
solutions with very small energies very efficiently, but it takes
an exponentially long time to actually reach the ground state. The
annealing approximation does not correctly describe very small
energies and cannot be used to establish its complexity. Note that
we can reconcile this with the fact that the annealing
approximation becomes \textsl{exact} in the limit when number of
bits in a clause $K \to \infty$: if the annealing approximation
fails for some $E \lesssim E_K$ we expect that $E_K$ is decreasing
to zero as $K$ increases. However for any finite $K$, the free
energy computed within the annealing approximation is free from
any singularities indicative of a phase transition. To study the
complexity of simulated annealing one needs to use the tools of
spin glass theory, in particular, the replica trick
\cite{SGTB,Monasson:prl96,Monasson:pre97} (see also below).

\begin{figure}[!ht]
\includegraphics[width=3.5in]{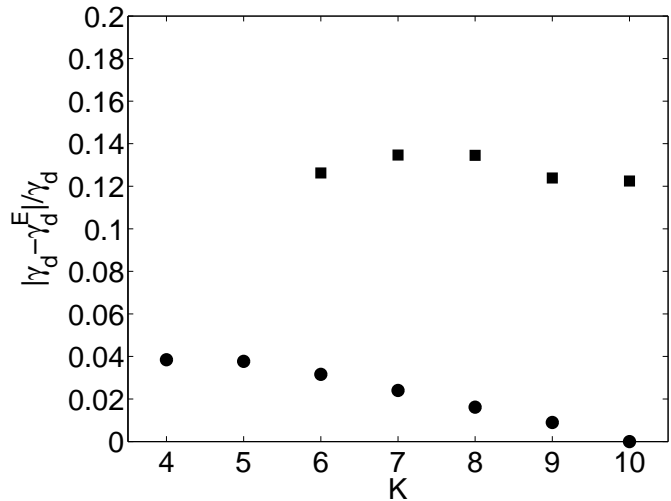}
\caption{\label{fig:error} Relative difference between predictions
for the dynamical phase transition point in the case of full
($\gamma_d$) and energy-only ($\gamma_{d}^{E}$)  landscapes  {\em
vs}  of $K$ for 1-in-K SAT (circles) and K-NAE SAT (squares).}
\end{figure}

In contrast, in our analysis of the quantum adiabatic algorithm,
we observed a first-order phase transition, and, importantly, it
happens for energies $E_* = O(E_\infty)$ (where $E_\infty$ is the
expected energy at infinite temperature, $E_\infty = \frac{1}{2^n}
\sum_z E_z$). Moreover, the energies on both sides of the
transition, relative to $E_\infty$ seem not to change appreciably
with increasing $K$. Since the annealing approximation for this
range of energies can be used, the prediction for the dynamic
transition should survive, though the exact numerical values may
acquire corrections. We have recomputed the dynamic transition
with simplified energy-only landscapes (see Fig. \ref{fig:error}).
For 1-in-K SAT one can clearly see that the relative correction
quickly diminishes. We believe that same happens for K-NAE SAT if
sufficiently large $K$'s are considered. If this indeed holds, it
serves as a corroboration that our results are correct numerically
for large $K$. It should be noted that the large-K limit
corresponds to the so-called random energy model, where one does
not expect to perform better then ${\cal O}(2^N)$ via any quantum
algorithm.

The idea of using energy-only landscapes was present in
\cite{Hogg:01} as well as \cite{Macready} and \cite{Smel}. A jump
in the time-dependence of the expected energy value was seen in
numerical simulations \cite{Hogg:02}, indicative of first-order
phase transition, though a different ensemble was considered (only
instances having a unique solution were considered).

We also attempted to go beyond simple annealing approximation and
studied the dynamical transition using its refinement. For that we
developed a polynomial mapping of the optimization problem defined
on a full graph onto the problem defined on its subgraph (a core)
where disorder-related fluctuations are significantly reduced and
annealing approximation is expected to perform much better. As a
test we used annealing approximation on a core to calculate the
position $\gamma_c$ of a static (satisfiability) transition where
the entropy of the state with $E$=0 vanishes. We  also computed
$\gamma_c$  numerically and found it to be very close to the
analytical result. We then studied the dynamics of quantum
adiabatic evolution algorithm on a core using an extended set of
landscape functions and found that old results obtained on a full
graph are reproduced qualitatively. This supports our earlier
prediction that the location of the phase transition is not very
sensitive to exact nature of annealing approximation employed.

We emphasize that different versions of annealing approximation
employed in this paper  describe the phase transition as a global
bifurcation between two macroscopic states (pure states) in the
space of macroscopic variables defined by a set of landscape
functions. The complexity is due to tunneling between the pure
states. In contrast, spin glass theory predicts the existence of
an infinite number of pure states \cite{SGTB} at sufficiently
small energies. On the other hand, as we mentioned above the
first-order quantum phase transition occurs for large energies
$E_*$ and this has been confirmed with an improved annealing
approximation. 

Although the transition is seemingly absent for small $K$, a
better approach (as compared to annealing approximation) may
reveal it. Moreover, we believe that if this happens the order of
the transition will remain unchanged, suggesting that the disorder
may be irrelevant for the determination of the order of the phase
transition and, consequently, for the complexity of the quantum
adiabatic algorithm. That is, the exponential complexity is not
due to the true combinatorial complexity of the underlying random
optimization problem but rather due to peculiarities of the driver
term and a particular ensemble of random instances considered.

A future extension of the  present work is to include sufficiently
large (possibly infinite) number of landscape parameters, thereby
making annealing approximation increasingly precise. In this
regard we recall  that 1-bit-flip conditional distribution over
landscape parameters  employed in this paper (\ref{Laplace}) can
be expressed via the set of coefficients $\{c_{\sigma,{\bf k}}\}$
that are concentrations of binary variables in a given string with
different types ($\sigma,{\bf k}$) of an immediate neighborhood.
In fact, these coefficients themselves can be used in an extended
set of landscape parameters ${\bf x}$. Then appropriate effective
potential $f({\bf x},\Gamma)$ can  be introduced and its
bifurcation can be studied when $\Gamma$ varies from $\infty$ to
$0$. Furthermore, one can consider introducing progressively larger
sets of landscape functions by defining neighborhoods  of
progressively larger size and using the well-known property that
local structure of random (hyper)graph is tree-like \cite{cavity}.

\section{Acknowledgments}
This work was supported in part by the National Security Agency
(NSA) and Advanced Research and Development Activity (ARDA) under
Army Research Office (ARO) contract number
ARDA-QC-P004-J132-Y03/LPS-FY2003, we also want to acknowledge the
partial support of NASA CICT/IS program.

We would also like to acknowledge helpful comments by E.~Farhi, 
J.~Goldstone (MIT) and S.~Gutmann (Northeastern U).

\appendix

\section{\label{app:np} On the NP-completeness of Positive $1$-in-$K$ Sat and
\newline Positive $K$-NAE-Sat.}

We set out to prove that both Positive $1$-in-$K$ Sat and
$K$-NAE-Sat are NP-complete. It is straightforward to see that it
takes a polynomial time to verify the assignment, so these problems
are in NP. We now prove that they are as hard as the Satisfiability
problem, which is NP-complete, by showing that any boolean formula
can be represented as an instance of these.

\subsection{Positive $1$-in-$K$ Sat}

A clause of type $( x, \ldots, x, y )$ necessarily implies $x = 0$
and $y = 1$; hence we can represent constants $0$ and $1$. A
clause of type $( 0, \ldots, 0, x, y )$ implies $x = \neg y$.
Finally, a clause of type $( 0, \ldots, 0, x, y, z )$ is equivalent
to a 3-clause $( x, y, z )$ so that we can restrict ourselves to
$K = 3$ without losing generality.

For $K = 3$, immediately observe that three clauses $( x, z, u' ) (
y, z, u'' ) ( u, u', u'' )$ with free variables $u, u', u''$
implies $z = \neg ( x \wedge y )$. This basic building block is in
fact sufficient to build any boolean formula, as a result, any
boolean formula can be cast as an $1$-in-$K$ SAT formula.

\subsection{Positive $K$-NAE-Sat}

A clause of type $( x, \ldots, x, y )$ necessarily implies $x =
\neg y$, and $( x, \ldots, x, y, z )$ is equivalent to $( x, y, z
)$ so we once again restrict ourselves to $K = 3$. In contrast to
$1$-in-$K$ problem, we shall require a non-trivial representation
of {\em false} or {\em true}. We will use {\em pairs} of variables
to denote variables of the boolean formula. Pairs $00$ or $11$
will represent value {\em false} and pairs $01$ or $10$ will
represent {\em true}.

The next building block, $( x, y, t ) ( y, z, t ) ( z, x, t )$
ensures that $t = 1$ if the majority of $x, y, z$ are $0$ and $t =
0$ if the majority are $1$. We shall use a shorthand $f ( t ; x,
y, z )$ to denote this. The expression $f ( z_1 ; x_1, y_1, y_2 )
f ( z_2 ; x_2, y_1, y_2 )$ then ensures $z = x \wedge y$ where $x,
y, z$ are represented as pairs $x_1 x_2$, $y_1 y_2$, $z_1 z_2$ as
indicated above. The operation of negation is trivial to
represent: if $x \equiv x_1 x_2$ then $\neg x \equiv ( \neg x_1 )
x_2$. These two are sufficient to construct any boolean formula.

\section{\label{app:next_order} Next order approximation for landscapes}

A better approximation for the values of critical
clause-to-variable rations can be obtained if we specify the
constraint that the distribution of vertex degrees be Poisson (as
it is supposed to be in a random hypergraph \cite{Mezard:02}). To
be precise, we specify that
\begin{equation}
 \sum_{\sigma,{\bf k}}
c_{\sigma,{\bf k}}  \prod_p \delta \left(
   \sum_{{\boldsymbol \alpha}} k_{{\boldsymbol \alpha}}^p - k_p \right) = c_{\{ k_p
   \}} \equiv \frac{\prod_p \gamma^{k_p}}{\prod_p k_p !} e^{- \gamma K} . \label{puasson}
   \end{equation}
   \noindent
Consequently, with this constraint the following expression for
$c_{\sigma,{\bf k}}$ is obtained: \begin{equation}
 c_{\sigma,{\bf
k}} = c_{\{ k_p \}}  \frac{e^{\lambda s} \prod_p \left[ k_p
   ! \prod_{{\boldsymbol \alpha}} ( \mu_{{\boldsymbol \alpha}}^p
   )^{k_{{\boldsymbol \alpha}}^p} / k_{{\boldsymbol \alpha}}^p ! \right]}{Z_{\{ k_p
   \}}}, \end{equation}
   \noindent
where we use
\begin{equation}
Z_{\{ k_p \}} = e^{\lambda} \prod_p \left( \sum_{\boldsymbol
\alpha} \delta(\alpha_p-1)\mu_{{\boldsymbol \alpha}}^p
\right)^{k_p} + e^{- \lambda} \prod_p \left( \sum_{\boldsymbol
\alpha} \delta(\alpha_p+1)\mu_{{\boldsymbol \alpha}}^p
\right)^{k_p}.\label{Zv}\end{equation} \noindent
 Annealed entropy
can be rewritten in the form
\begin{equation} 
   s_{\rm ann} [ q, \{ {\cal M}_{{\boldsymbol \alpha}} \} ] = \min_{\lambda,
   \mu_{{\boldsymbol \alpha}}^p} \left\{ - \lambda q + \sum_{p,{\boldsymbol \alpha}}
   {\cal M}_{{\boldsymbol \alpha}} \ln
   \frac{{\cal M}_{{\boldsymbol \alpha}}}{\mu_{{\boldsymbol \alpha}}^p} + \ln Z [ \lambda,
   \mu_{{\boldsymbol \alpha}}^p ] \right\} - \sum_{{\boldsymbol \alpha}}
   {\cal M}_{{\boldsymbol \alpha}} \ln {\cal M}_{{\boldsymbol \alpha}}+ \gamma \ln \gamma  - \gamma K, 
\end{equation}
where $\ln Z$ is given by
\begin{equation} 
\ln Z = \sum_{\{ k_p \}} c_{\{ k_p \}} \ln Z_{\{ k_p \}} . 
\end{equation}
The equations relating $q, \{ {\cal M}_{{\boldsymbol \alpha}} \}$ and
$\lambda, \{ \mu_{{\boldsymbol \alpha}}^p \}$ are given in
(\ref{S1}),(\ref{M1}).

 Similarly to Sec.~\ref{sec:potential} we
will use the notation (\ref{notation}). Since $\ln Z$ depends only
on $\lambda$ and $\mu_{\pm}$, $\partial \ln Z /
\partial \mu_{{\boldsymbol \alpha}}^p$ depends only on $\sigma_p$. Therefore,
${\cal M}_{{\boldsymbol \alpha}} / \mu_{{\boldsymbol \alpha}}^p =
{\cal M}_{\sigma_p} / \mu_{\sigma_p}$. Correspondingly,
\begin{eqnarray}
 s_{\rm ann} [ q, \{ {\cal M}_{{\boldsymbol \alpha}} \} ] & = & \min_{\lambda,
   \mu_{\pm}} \left\{ - \lambda q + {\cal M}_+ \ln \frac{{\cal M}_+}{\mu_+} + {\cal M}_- \ln
   \frac{{\cal M}_-}{\mu_-} + \ln Z \right\} \nonumber \\ && - \sum_{{\boldsymbol \alpha}}
   {\cal M}_{{\boldsymbol \alpha}} \ln {\cal M}_{{\boldsymbol \alpha}}+ \gamma \ln \gamma  - \gamma K 
\end{eqnarray}
For convenience, we introduce new variables
\begin{equation}
\mu_{\pm} = \mu \,e^{\pm h}.\label{muh}
\end{equation}
\noindent We then readily obtain
\begin{equation} \ln Z = \gamma \ln \mu + \sum_k c_k \ln [ 2 \cosh ( \lambda + k h ) ], \end{equation}
(where $k = \sum_p k_p$ and $c_k = \frac{( \gamma K )^k}{k!} e^{-
\gamma K}$) and $\mu$ drops out of the expression for $s_{\rm ann}$
altogether:
\begin{eqnarray}
 s_{\rm ann}[ q,\{{\cal M}_{\boldsymbol\alpha}\}] &=& \min_{\lambda, h} \left\{ -
\lambda q -
   ( {\cal M}_+ - {\cal M}_- ) h + \sum_k c_k \ln [ 2 \cosh ( \lambda + kh ) ]
   \right\}\\\nonumber
    &-& \sum_{{\boldsymbol \alpha}} {\cal M}_{{\boldsymbol \alpha}} \ln
   {\cal M}_{{\boldsymbol \alpha}} + {\cal M}_+ \ln {\cal M}_+ + {\cal M}_- \ln {\cal M}_{-}+ \gamma \ln \gamma - \gamma K. \label{Hannv}
\end{eqnarray}
\noindent
   It is easy to see from this expression what the equations for
$\lambda, h$ are:
\begin{eqnarray}
  \sum_k c_k \tanh [ \lambda + k h ] & = & q,\label{lambdanew}\\
  \sum_k kc_k \tanh [ \lambda + k h ] & = & {\cal M}_+ - {\cal M}_-
 .\nonumber
\end{eqnarray}
\noindent We now turn our attention to the function
$\ell(q,\{{\cal M}_{\boldsymbol\alpha}\})$ given by (\ref{Laplace}) with
$\theta$ and $y_{\boldsymbol\alpha}$ evaluated from
Eqs.~(\ref{theta}),(\ref{relationS}). The computation of
$e^{y_{{\boldsymbol \alpha}} - y_{{\boldsymbol \alpha}'}}$ yields
\begin{equation}
e^{y_{{\boldsymbol \alpha}} - y_{{\boldsymbol \alpha}'}} = \left(
   \sqrt{\frac{{\cal M}_+}{{\cal M}_-}} \sqrt{\frac{\mu_-}{\mu_+}} \right)^{\sigma_p}
   \sqrt{\frac{{\cal M}_{{\boldsymbol \alpha}'}}{{\cal M}_{{\boldsymbol \alpha}}}}
   .\label{yy}
   \end{equation}\noindent
Multiplied by $\mu_{{\boldsymbol \alpha}}^p$ this becomes
\begin{equation}
\mu_{{\boldsymbol \alpha}}^p \,e^{y_{{\boldsymbol \alpha}} -
   y_{{\boldsymbol \alpha}'}} = \frac{\sqrt{{\cal M}_{{\boldsymbol \alpha}}
   {\cal M}_{{\boldsymbol \alpha}'}}}{\sqrt{{\cal M}_+ {\cal M}_-}} \mu .
\end{equation}
   \noindent
The expression for $\ell(q,\{{\cal M}_{\boldsymbol\alpha}\})$ can be
written in the form (cf. (\ref{ell0}))
\begin{equation} \ell(q,\{{\cal M}_{\boldsymbol\alpha}\}) = \sum_{\{ k_p \}} \frac{c_{\{ k_p \}}}{Z_{\{ k_p \}}}
   \sum_{\sigma,{\bf k}}^{\prime} e^{\lambda s + 2 xs} \prod_p \left[ k_p !
   \prod_{{\boldsymbol \alpha}} ( \mu_{{\boldsymbol \alpha}}^p
   e^{y_{{\boldsymbol \alpha}} - y_{{\boldsymbol \alpha}'}} )^{k_{\alpha}^p} /
   k_{\alpha}^p ! \right], 
\end{equation}
with the internal sum running over ${\bf k}$ consistent with a set of
$\{ k_p \}$ (\ref{puasson}). Substituting the quantities defined
above this becomes
\begin{equation} \ell(q,\{{\cal M}_{\boldsymbol\alpha}\}) = \sum_{\{ k_p \}} c_{\{ k_p \}}  \frac{\sum_{\sigma} \prod_p
\left( \mu\,
   \sum_{\boldsymbol \alpha}
\delta(\alpha_p-\sigma)   \frac{\sqrt{{\cal M}_{{\boldsymbol \alpha}}
{\cal M}_{{\boldsymbol \alpha}'}}}{\sqrt{{\cal M}_+ {\cal M}_-}}
   \right)^{k_p}}{Z_{\{ k_p \}}}.\label{ellv1}
\end{equation}
\noindent After some transformations we finally obtain
\begin{equation}\ell(q,\{{\cal M}_{\boldsymbol\alpha}\})    = \sum_k c_k
\frac{\left(
   \frac{\sum_{<{\boldsymbol \alpha},{\boldsymbol \alpha}' >}
   \sqrt{{\cal M}_{{\boldsymbol \alpha}} {\cal M}_{{\boldsymbol \alpha}'}}}{\sqrt{{\cal M}_+ {\cal M}_-}}
   \right)^k}{\cosh [ \lambda + kh ]} . \label{ellv2}
   \end{equation}
   \noindent
where $\lambda,\,h$ are given by
(\ref{notation}),(\ref{lambdanew}). Using symmetric ansatz
(\ref{ansatz}) it is straightforward to calculate from
(\ref{ellv2}) the restricted function $\bar \ell(q,\{{\cal
M}_m\})$ (cf. (\ref{ell2})). We must note however, that although
this represents a next-order improvement over annealing
approximation, the relative changes in $\gamma_c$ and $\gamma_d$
computed with this improved approximation are nearly imperceptible
($\sim$ 10$^{-4}$).

\end{document}